\documentclass[11pt]{article}
\usepackage[margin=2.5cm]{geometry}

\usepackage[version=3]{mhchem}
\usepackage{bm}
\usepackage{siunitx}
\usepackage{booktabs}
\usepackage{amsmath}
\usepackage{graphicx}
\usepackage[super,sort&compress,comma]{natbib}
\usepackage{hyperref}
\providecommand*{\doi}[1]{\href{https://doi.org/#1}{\nolinkurl{#1}}}

\newcommand*{\rAct}{\rho^{\mathrm{act}}}
\newcommand*{\rEnv}{\rho^{\mathrm{env}}}
\newcommand*{\rTot}{\rho^{\mathrm{tot}}}
\newcommand*{\rEnvF}{\tilde{\rho}^{\mathrm{env}}}
\newcommand*{\vemb}{v_{\mathrm{emb}}}
\newcommand*{\Eint}{E^{\mathrm{int}}}
\newcommand*{\naddxc}{E_{\mathrm{xc}}^{\mathrm{nadd}}}
\newcommand*{\naddke}{T_{\mathrm{s}}^{\mathrm{nadd}}}
\newcommand*{\mR}{\mathbf{R}}
\newcommand*{\mr}{\mathbf{r}}
\newcommand*{\mL}{\mathbf{L}}
\newcommand*{\matD}{\mathbf{D}}
\newcommand*{\matS}{\mathbf{S}}
\newcommand*{\matW}{\mathbf{W}}
\newcommand*{\matV}{\mathbf{V}}
\newcommand*{\Tr}{\mathrm{Tr}}

\title{Quantum Embedding from Molecular to Crystalline Environments at the Cost of a Gas-Phase Calculation}
\author{Manas Sharma$^{1,2}$ \and Marek Sierka$^{1,*}$}
\date{}

\begin{document}

\maketitle

\noindent $^{1}$Otto Schott Institute of Materials Research, Friedrich Schiller University Jena, L\"obdergraben 32, 07743 Jena, Germany\\
$^{2}$Department of Chemical Engineering, Indian Institute of Science, Bengaluru, Karnataka, 560012, India\\
$^{*}$Corresponding author: marek.sierka@uni-jena.de

\begin{abstract}
We present a direct-space formulation and implementation of quantum density functional theory (DFT) embedding for molecular active subsystems embedded in frozen environments (hosts) of arbitrary dimensionality, from finite molecular hosts to extended systems. 
The host is computed once without the embedded molecule (the guest) and is represented only by its electron density fitted in an atom-centered auxiliary Gaussian basis together with its nuclear charges with the periodic electrostatics separated into an explicit near-field and a continuous fast-multipole far-field operator.
The method allows the removal of host atoms from this representation to create vacancies or other voids without requiring a supercell creation. 
Analytic nuclear gradients are also implemented, enabling structural relaxation of the active subsystem.
We demonstrate the versatility of the approach through applications to adsorbates on periodic surfaces and molecules embedded in bulk solvent environments.
For CO on MgO(001), we show that DFT-in-DFT converges rapidly and reproduces the periodic interaction energy within 0.44~kJ/mol while reducing the cost by a factor of 27. 
Real-time time-dependent embedded DFT (RT-TDDFET) of methylenecyclopropene in periodic water is showcased to produce a $\sim$0.6~eV solvatochromic blue shift, while strong-field propagation of CO$_2$ on hexagonal boron nitride (hBN) yields substrate-induced effects at the cost of a gas-phase propagation. 
Coupling the embedding scheme to excited-state correlated wave function theory methods such as CC2 yields a solvent shift within 0.01~eV of supermolecular CC2 references, while reducing the computational cost by over 500 times.
\end{abstract}

\section{Introduction}

Hybrid chemical systems of practical interest consist of a localized region, which
determines the property of interest, and an extended environment, which modifies it.
Typical examples are molecules adsorbed on surfaces, chromophores in solution or in
molecular crystals, and point defects in solids. An accurate description of the local
region often requires computationally demanding approaches, such as hybrid density
functionals, correlated wave function theory (WFT) methods, or an explicit treatment of
the electron dynamics, whereas the environment acts on this region mainly through
electrostatic interactions, Pauli repulsion, and polarization. Treating the complete
system at the level required for the local region is therefore usually prohibitive. For
extended environments, the periodic boundary conditions introduce a further difficulty,
since an adsorbate or a defect is repeated in every cell, and supercells that are large
enough to suppress the spurious interactions between its periodic images can easily
contain hundreds of atoms.\cite{Freysoldt.Grabowski.ea:First-principles.2014} Moreover,
many applications require not a single calculation but a large number of them, for
example, for scans of potential energy surfaces, structure optimizations, the screening
of adsorption sites, or molecular dynamics. Embedded-cluster and hybrid quantum
mechanics/quantum mechanics approaches, which combine a high-level treatment of a finite
cluster with a lower-level description of the extended system, allow reaching chemical
accuracy for molecule--surface interactions.\cite{Sauer:Ab.2019} More generally, quantum
embedding methods partition the total system into an active subsystem and an
environment, which are described quantum mechanically at different levels of
theory.\cite{Sun.Chan:Quantum.2016,Jones.Mosquera.ea:Embedding.2020,%
Libisch.Huang.ea:Embedded.2014,Jacob.Neugebauer:Subsystem.2014,%
Jacob.Neugebauer:Subsystem-Update.2024,Wesolowski.Shedge.ea:Frozen-Density.2015}

Density functional theory (DFT) based quantum embedding uses the electron density to define this
partition. Its most common realization is frozen density embedding
(FDE),\cite{Senatore.Subbaswamy:Density.1986,Cortona:Self-consistently.1991,%
Wesolowski.Warshel:Frozen.1993} in which the density of the active subsystem is
optimized in the presence of a fixed environment density. The environment enters through
an embedding potential that consists of its electrostatic potential and of nonadditive
exchange-correlation and kinetic contributions. Since the exact nonadditive kinetic
potential is unknown, it is evaluated with an approximate kinetic energy density functional
(KEDF).\cite{Gotz.Beyhan.ea:Performance.2009} Projection-based embedding avoids KEDFs
by enforcing the orthogonality of the occupied orbitals of the
subsystems,\cite{Manby.Stella.ea:Simple.2012,Huzinaga.Cantu:Theory.2003,%
Chulhai.Jensen:Frozen.2015,Lee.Welborn.ea:Projection-Based.2019} but requires that each subsystem is expanded in the
basis of the full system, so that the cost of the embedded calculation remains tied to
the size of the full system. 
Both families of methods allow the active subsystem to be described with WFT methods, leading to WFT-in-DFT
embedding,\cite{Govind.Wang.ea:Accurate.1998,Govind.Wang.ea:Electronic-structure.1999,%
Gomes.Jacob.ea:Calculation.2008,Libisch.Huang.ea:Embedded.2014,%
Lee.Welborn.ea:Projection-Based.2019} application to excited
states,\cite{Gomes.Jacob.ea:Calculation.2008,Daday.König.ea:State-Specific.2013,%
Daday.Konig.ea:Wavefunction.2014} and coupling with real-time time-dependent
DFT (RT-TDDFT).\cite{Krishtal.Ceresoli.ea:Subsystem.2015,%
De-Santis.Belpassi.ea:Environmental.2020,Sharma.Sierka:Efficient.2022}

Periodic environments are indispensable for adsorbates on surfaces, molecules in crystals
and liquids, and defects in solids, and several approaches have extended DFT-based
embedding to them. Carter and co-workers pioneered the embedding of correlated wave
functions in periodic DFT environments,\cite{Govind.Wang.ea:Accurate.1998,%
Govind.Wang.ea:Electronic-structure.1999,Kluner.Govind.ea:Prediction.2001,%
Kluner.Govind.ea:Periodic.2002,Huang.Pavone.ea:Quantum.2011,%
Libisch.Huang.ea:Embedded.2014,Yu.Libisch.ea:Implementation.2015} combining a plane-wave
description of the periodic system with localized basis functions for the embedded
cluster. The embedding potential then has to be transferred between the two
representations, and in density functional embedding theory it is obtained by a
potential inversion that requires the density of the full
system.\cite{Huang.Pavone.ea:Quantum.2011} Periodic subsystem DFT has been implemented in
plane-wave codes,\cite{Genova.Ceresoli.ea:Periodic.2014,Genova.Ceresoli.ea:eQE.2017}
including a real-time variant,\cite{Krishtal.Ceresoli.ea:Subsystem.2015} and in the
Gaussian and plane-wave framework of CP2K, where it has been employed for condensed-phase
molecular dynamics,\cite{Iannuzzi.Kirchner.ea:Density.2006,%
Andermatt.Cha.ea:Combining.2016} for nonadiabatic dynamics,\cite{Malis.Luber:SCF.2021}
and for the embedding of complete active space wave functions in periodic DFT
environments.\cite{Schreder.Luber:Implementation.2024} A density functional embedding
theory based on products of basis functions has been formulated within the same
framework.\cite{Rybkin:Formulation.2021} H{\"o}fener and co-workers combined wave function
FDE with one-dimensional periodicity\cite{Fink.Hofener:Combining.2021,%
Wachter-Lehn.Fink.ea:Wavefunction.2022} and treated local perturbations in
two-dimensional periodic molecular environments.\cite{Weissert.Ellinger.ea:Employing.2025}
For projection-based embedding, Chulhai and Goodpaster introduced a periodic formulation
based on PySCF that includes $\mathbf{k}$-point sampling and treats the active subsystem
either periodically or as a molecular cluster.\cite{Chulhai.Goodpaster:Projection-Based.2018}
As in the molecular case, this scheme is exact only in the limit in which each subsystem
is described with the basis of the full periodic
system.\cite{Chulhai.Goodpaster:Improved.2017,Chulhai.Goodpaster:Projection-Based.2018}
Regional embedding\cite{Lau.Knizia.ea:Regional.2021} and a recent interface between CP2K
and Q-Chem\cite{Fonlon.Kolodzeiski.ea:Q-Chem.2026} select the active orbitals from a
periodic mean-field calculation of the full system. Hence, in most of these approaches,
the periodic environment and the molecular active subsystem are described in different
representations, a calculation of the full system is required, or the active subsystem
itself is subject to the periodic boundary conditions.

In our previous work~\cite{Sharma.Sierka:Efficient.2022}, we reported an implementation of DFT-based embedding in the TURBOMOLE
program package,\cite{Franzke.Holzer.ea:TURBOMOLE.2023,Franzke2026RecentTurbomole} in
which molecular and periodic systems are both described with Gaussian-type orbitals. It
builds on the periodic DFT module \textsc{riper},\cite{Burow.Sierka.ea:Resolution.2009,%
Burow.Sierka:Linear.2011,Lazarski.Burow.ea:Density.2015,Lazarski.Burow.ea:Density.2016,%
Grajciar:Low.2015,Becker.Sierka:Density.2019,Sharma.Franzke.ea:Density.2025} which
evaluates the Kohn-Sham (KS) matrix of periodic systems entirely in direct space by
combining density fitting (DF) with the continuous fast multipole method
(CFMM).\cite{White.Johnson.ea:continuous.1994,Lazarski.Burow.ea:Density.2015} Since crystal
orbitals, molecular orbitals, and electron densities share the same atom-centered
representation, the embedding potential of a periodic environment can be formed directly
in the basis of a molecular active subsystem, without any transformation between plane
waves and localized functions. The implementation has been coupled to correlated WFT and
Green's function methods for adsorption energies, excitation energies of solvated
molecules, and optical gaps of ionic solids,\cite{Sharma.Sierka:Efficient.2022,%
Sharma.Sierka:Optical.2024} and to RT-TDDFT, leading to real-time time-dependent density
functional embedding theory (RT-TDDFET).\cite{Mueller.Sharma.ea:Real-time.2020,%
Sharma.Sierka:Efficient.2022}

For molecules in periodic environments, however, this implementation followed a
subtractive strategy related to the early work of Carter and
co-workers,\cite{Kluner.Govind.ea:Prediction.2001,Kluner.Govind.ea:Periodic.2002} which
entailed four significant limitations. First, the environment density was not obtained
from a calculation of the environment itself but approximated as the difference between
the density of the combined periodic system and that of the isolated active subsystem.
Such a difference density is not guaranteed to be nonnegative. Second, the embedding
potential was constructed once from these densities and kept fixed throughout the
self-consistent field (SCF) optimization of the active subsystem, since updating it with
the embedded active density caused convergence
problems.\cite{Kluner.Govind.ea:Periodic.2002,Lahav.Kluner:self-consistent.2007,%
Sharma.Sierka:Efficient.2022}
Third, a periodic calculation of the
combined system was required for every structure of the active subsystem, and a point
defect required a periodic calculation of the defective
supercell.\cite{Sharma.Sierka:Optical.2024} The most expensive step was thus never
avoided, and embedding reduced only the cost of the high-level treatment of a given
structure. Fourth, analytic nuclear gradients were not available, which precluded
structure relaxations of the embedded subsystem. Together, these limitations confined the
earlier implementation to single-point calculations at fixed geometries.

In this work, we introduce a formulation of DFT-based embedding of molecules in extended
environments that removes all four limitations. In \textsc{riper}, the electron density of a periodic system is expanded in
atom-centered auxiliary Gaussian functions by Coulomb-metric density
fitting,\cite{Burow.Sierka.ea:Resolution.2009,Lazarski.Burow.ea:Density.2015} and this
fitted density is the source of the Coulomb potential in the periodic calculation
itself. We use the fitted density of the clean host as the complete and only
representation of the frozen environment electrons. The embedded calculation thus
receives the fitting coefficients, the auxiliary basis set, the nuclear positions and
charges, and the lattice vectors from a single calculation of the host, and nothing else.
No orbitals, density matrices, or $\mathbf{k}$-point dependent quantities of the host are
needed, and no calculation of the combined system is performed at any stage.
Throughout this work, the host is the clean molecular or periodic system that is
calculated once, for example a substrate or a solvent. The guest is the molecule of
interest, for example an adsorbate or a solute. The active subsystem is the part
whose density is optimized in the embedded calculation: the guest together with any
host atoms promoted to it.
The environment is the frozen part of the host that the active subsystem experiences.
Without a void, it is the complete host. With a void, created by deleting selected host
atoms from the host density, it is the remainder of the host. A void serves two
purposes. It makes room for promoted host atoms, such as the substrate atoms around an
adsorption site, and it models defects, such as vacancies and substitutions.

\begin{figure}
  \centering
  \includegraphics[width=\linewidth]{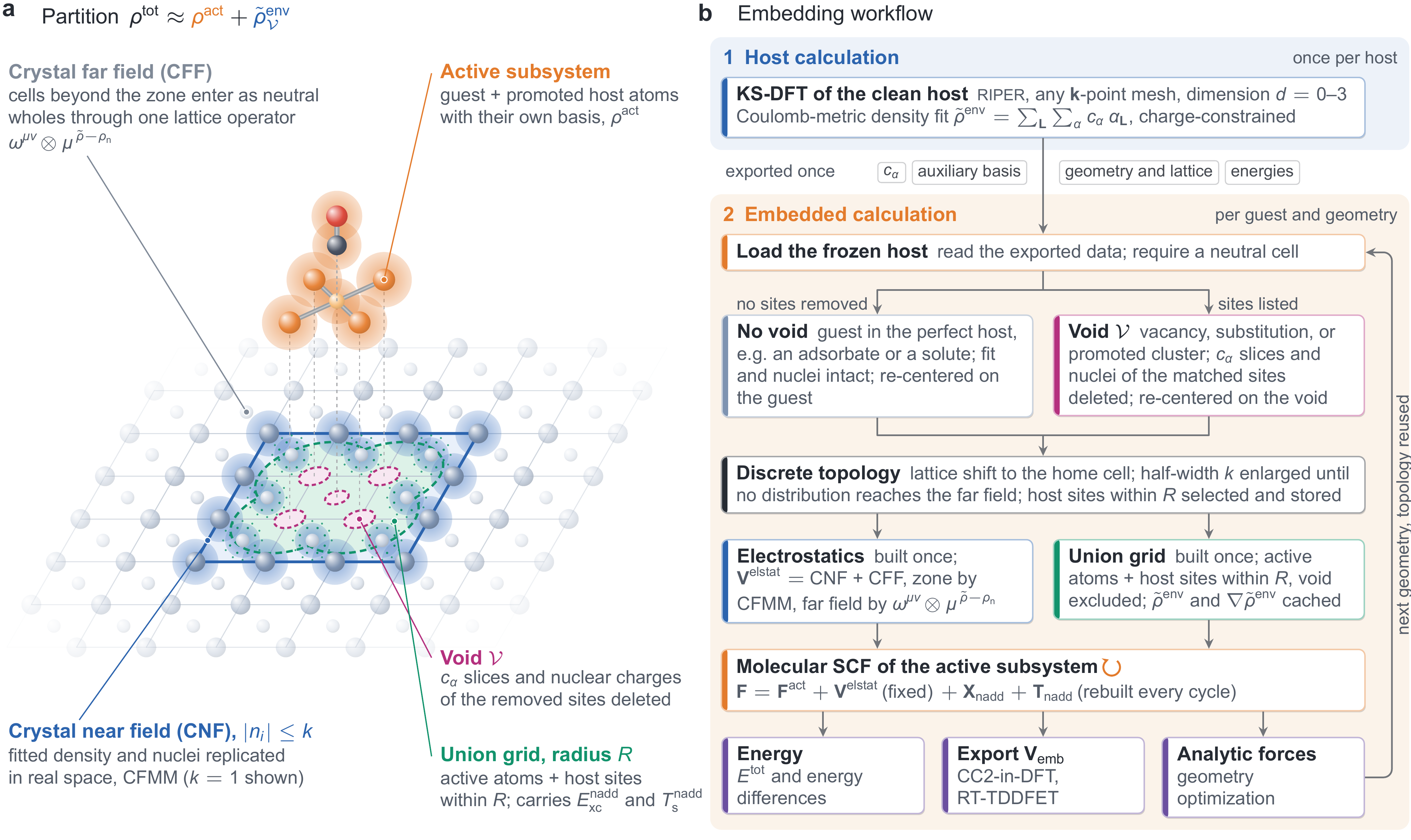}
  \caption{Direct-space embedding of a guest molecule in a periodic host. \textbf{a}, Partition of the
  construction, drawn for a two-dimensional host with an explicit zone of half width
  $k=1$. The frozen host is represented only by its fitted auxiliary density (blue halos)
  and nuclear charges. The cells of the crystal near field (CNF, blue frame) are
  replicated explicitly in real space, and every cell beyond it enters as a neutral whole
  through the crystal far-field (CFF) lattice operator. In a void calculation the
  promoted host atoms are deleted from the fit and from the nuclear charges (dashed
  magenta sites) and rejoin the guest, here CO, in the active subsystem with their own
  basis functions. The active subsystem is lifted for clarity. The dashed green line
  encloses the union quadrature grid, which spans the active atoms and the host sites
  within the selection radius $R$ (green points) and carries the nonadditive
  exchange-correlation and kinetic terms. \textbf{b}, Embedding workflow. The host calculation
  runs once and exports the fit coefficients, a manifest, and the host energies. Each
  embedded calculation verifies these data, removes the listed sites when a void is
  present, fixes the discrete topology, builds the electrostatic matrix and the cached
  host density once, and runs a molecular SCF in which only the nonadditive terms are
  rebuilt. The converged embedding potential is exported for WFT-in-DFT and RT-TDDFET
  calculations, and analytic forces drive structure optimizations that reuse the stored
  topology. The \texttt{riperembed} driver automates both stages.}
  \label{fig:scheme}
\end{figure}

This interface has several consequences. First, the environment
density is the fitted ground-state density of the host instead of a difference of two
densities. Its electrostatic potential reduces to three-center integrals and lattice
sums, which are evaluated with an explicit near field and a CFMM far field, that is, with
the same machinery as in the host calculation, and a single construction covers
molecular hosts as well as periodic hosts of any dimensionality. Second, the nonadditive
exchange-correlation and kinetic potentials are evaluated from the same fitted density
on a quadrature grid and are rebuilt in every SCF iteration, so that the KS equations
with constrained electron density are solved self-consistently for the frozen host.
Third, since the fit is partitioned over atoms, deleting the contributions of selected host atoms carves a well-defined void into the host (Figure~\ref{fig:scheme}a), whose remainder then acts as the environment. 
The active subsystem can thereby replace an
arbitrary part of the host, so that adsorbates together with host atoms promoted to the active subsystem,
vacancies, substitutions, or a molecule replacing its own motif in a molecular crystal
can be treated without a calculation of a defective supercell. Because the active
subsystem is a finite, nonperiodic system, the embedded calculation describes an isolated guest in
an otherwise perfect, infinite host, free of interactions between periodic images of the
guest. Fourth, once the host has been generated, an embedded calculation costs about as
much as a gas-phase calculation of the active subsystem, and the same host can be reused
for any number of guests and structures. Since the embedded calculation is an ordinary
molecular SCF calculation, the full functionality of the molecular code, including
hybrid and range-separated hybrid functionals, is available for the active subsystem.
Finally, the scheme is agnostic about the origin of the host density, since any method
that provides a density fitted in atom-centered Gaussian functions can serve as the
source. The approximate nonadditive kinetic potential remains the main approximation of
the scheme. Its effect can, however, be reduced systematically by promoting host atoms
near the guest to the active subsystem, whose cost grows only with the size of this local
region.

Because the embedding potential is now the functional derivative of the embedded energy,
the energy is stationary with respect to the converged active density, and analytic
nuclear gradients can be derived along the usual lines. We derive and implement the analytic gradient of the complete embedding
functional with respect to the coordinates of the active nuclei. It contains the
derivatives of the near-field and far-field electrostatic interactions with the host and,
for the nonadditive terms, contributions from the motion of the quadrature points through
the frozen host density and from the derivatives of the quadrature weights. All terms are
validated against finite differences (Supplementary Note~1, Supplementary Tables~1 and~2). Analytic gradients have been reported for FDE in
molecular environments\cite{Schluns.Franchini.ea:Analytical.2017,%
Heuser.Hofener:Wave-function.2016,Klahr.Schluns.ea:Geometry.2018} and are available in
periodic subsystem DFT schemes in which all subsystems are
periodic.\cite{Iannuzzi.Kirchner.ea:Density.2006,Andermatt.Cha.ea:Combining.2016} To our
knowledge, however, they have not been reported for a molecular subsystem in a frozen
periodic host whose electrostatics is represented by lattice-summed multipole operators.
Since the frozen host can exert a net force and torque on the active subsystem, the
rigid-body components of this gradient carry physical meaning and must not be projected
out. Beyond DFT, the converged embedding potential is exported as a one-electron matrix
and enters WFT-in-DFT and RT-TDDFET calculations without
modification.\cite{Sharma.Sierka:Efficient.2022} As no calculation of the combined system
is required, correlated and time-dependent calculations of molecules in extended
environments become available at essentially the cost of the corresponding gas-phase
calculations. Taken together, these developments extend DFT-based embedding of molecules
in extended environments from single-point calculations at fixed geometries to potential
energy scans, structure relaxations, and electron dynamics, at a cost that is determined
by the size of the active subsystem rather than by that of the environment.

The accuracy and efficiency of the approach are assessed for adsorbates on surfaces and
for molecules in bulk solvent. For CO on MgO(001), the interaction energy curve from
DFT-in-DFT embedding is compared with periodic calculations of the combined system, and
its convergence with the number of promoted host atoms is analyzed together with the
computational cost. Structure relaxations of NH$_3$ on MgO(001) with the analytic
gradients test the accuracy of embedded adsorption energies and geometries. Excited
states are addressed by RT-TDDFET calculations of the solvatochromic shift of
methylenecyclopropene (MCP) in periodic water, which are compared with linear-response
time-dependent DFT (LR-TDDFT) calculations of the complete periodic system and with the
earlier subtractive embedding potential. The response of an adsorbate to a strong laser field is examined
through the high harmonic generation (HHG) spectrum of CO$_2$ on a hexagonal boron
nitride (hBN) monolayer. Finally, excitation energies of MCP in periodic water from the
second-order approximate coupled cluster singles and doubles model
(CC2)\cite{Christiansen.Koch.ea:second-order.1995,Hattig.Weigend:CC2.2000} embedded in
DFT (CC2-in-DFT) are compared with supermolecular CC2 calculations of clusters that
contain complete hydration shells.

\section{Results}
\label{sec:results}

\subsection{Frozen density embedding}
\label{sec:fde}

We consider the closed-shell case. The open-shell extension follows the usual FDE
formalism. Within FDE,\cite{Wesolowski.Warshel:Frozen.1993} the total electron density is
partitioned as
\begin{equation}
  \rTot(\mr) = \rAct(\mr) + \rEnv(\mr),
  \label{eq:partition}
\end{equation}
and the total energy is a bifunctional of the two subsystem densities,
\begin{equation}
  E^{\mathrm{tot}}[\rAct,\rEnv]
  = E^{\mathrm{act}}[\rAct] + E^{\mathrm{env}}[\rEnv] + \Eint[\rAct,\rEnv],
  \label{eq:etot}
\end{equation}
where $E^{\mathrm{act}}$ and $E^{\mathrm{env}}$ are the KS-DFT energies of
the isolated subsystems. The interaction energy reads
\begin{align}
  \Eint[\rAct,\rEnv] ={}&
  \left(\rAct\,\middle|\,\rEnv - \rho_{\mathrm{n}}^{\mathrm{env}}\right)
  - \left(\rho_{\mathrm{n}}^{\mathrm{act}}\,\middle|\,\rEnv\right)
  + \left(\rho_{\mathrm{n}}^{\mathrm{act}}\,\middle|\,
    \rho_{\mathrm{n}}^{\mathrm{env}}\right) \nonumber\\
  &+ \naddxc[\rAct,\rEnv] + \naddke[\rAct,\rEnv],
  \label{eq:eint}
\end{align}
where $\rho_{\mathrm{n}}^{\mathrm{act}}$ and $\rho_{\mathrm{n}}^{\mathrm{env}}$
are the positive nuclear charge densities of the two subsystems. Round brackets denote
Coulomb integrals, $(f|g) = \iint f(\mr)\,g(\mr')\,|\mr-\mr'|^{-1}\,
\mathrm{d}\mr\,\mathrm{d}\mr'$. The nonadditive exchange-correlation and kinetic
energies are
\begin{align}
  \naddxc[\rAct,\rEnv] &= E_{\mathrm{xc}}[\rAct+\rEnv] - E_{\mathrm{xc}}[\rAct]
  - E_{\mathrm{xc}}[\rEnv],
  \label{eq:naddxc}\\
  \naddke[\rAct,\rEnv] &= T_{\mathrm{s}}[\rAct+\rEnv] - T_{\mathrm{s}}[\rAct]
  - T_{\mathrm{s}}[\rEnv].
  \label{eq:naddke}
\end{align}
Here $E_{\mathrm{xc}}$ is the exchange-correlation and $T_{\mathrm{s}}$ the noninteracting kinetic energy functional. For a given frozen, nonnegative $\rEnv$, minimization of eq~\ref{eq:etot} with respect
to $\rAct$ leads to KS equations with constrained electron density (KSCED),\cite{Wesolowski.Warshel:Frozen.1993,Wesolowski.Weber:Kohn-Sham.1996} in which the
KS potential of the isolated active subsystem is augmented by the embedding potential
\begin{equation}
  \vemb[\rAct,\rEnv](\mr)
  = \int \frac{\rEnv(\mr') - \rho_{\mathrm{n}}^{\mathrm{env}}(\mr')}{|\mr-\mr'|}\,
  \mathrm{d}\mr'
  + \frac{\delta \naddxc}{\delta \rAct(\mr)}
  + \frac{\delta \naddke}{\delta \rAct(\mr)},
  \label{eq:vemb}
\end{equation}
whose first term is the electrostatic potential of the environment electrons and nuclei.
The nonadditive kinetic potential is evaluated with an approximate
KEDF.\cite{Lembarki.Chermette:Obtaining.1994,Gotz.Beyhan.ea:Performance.2009} 
We denote this scheme KEDF-based embedding.

In the matrix representation over the atomic orbitals $\mu,\nu$ of the active
subsystem, the embedding contribution to the KS matrix is
\begin{equation}
  \matV_{\mathrm{emb}}
  = \matV^{\mathrm{elstat}}
  + \mathbf{X}_{\mathrm{nadd}} + \mathbf{T}_{\mathrm{nadd}}.
  \label{eq:vembmat}
\end{equation}
Here $\mathbf{X}_{\mathrm{nadd}}$ and $\mathbf{T}_{\mathrm{nadd}}$ are the matrix
representations of the nonadditive exchange-correlation and kinetic potentials of
eq~\ref{eq:vemb}. The electrostatic matrix $\matV^{\mathrm{elstat}}$ represents the first term of
eq~\ref{eq:vemb}. It is independent of $\rAct$ and is constructed once for an embedded
calculation. The nonadditive terms depend on $\rAct$ and are rebuilt in every
SCF iteration. Two basis conventions are
supported.\cite{Sharma.Sierka:Efficient.2022} In a monomolecular basis, each subsystem
uses basis functions on its own atoms. In a supersystem basis, both subsystems use basis
functions on all atoms. 
The equations below first use the monomolecular convention. 
We state the additional supersystem terms where they enter.

\subsection{Direct-space embedding potential of a periodic environment}
\label{sec:periodic}

The frozen environment density is obtained from an individual periodic KS-DFT calculation of the clean host with \textsc{riper} (stage 1 in Figure~\ref{fig:scheme}b).\cite{Sharma.Franzke.ea:Density.2025,Lazarski.Burow.ea:Density.2015,%
Lazarski.Burow.ea:Density.2016} 
We use the notation of
ref~\citenum{Lazarski.Burow.ea:Density.2015} in the following. 
In our density fitting scheme,\cite{Burow.Sierka.ea:Resolution.2009,Lazarski.Burow.ea:Density.2015} the host density is approximated by an auxiliary crystal density
\begin{equation}
  \rEnv(\mr) \approx \rEnvF(\mr) = \sum_{\mL} \tilde{\rho}_{\mL}(\mr),
  \qquad
  \tilde{\rho}_{\mL} = \sum_{\alpha} c_\alpha\, \alpha_{\mL},
  \label{eq:rienv}
\end{equation}
where $\alpha_{\mL}$ is an atom-centered auxiliary Gaussian-type orbital in cell $\mL$.
The coefficients $c_\alpha$ do not depend on $\mL$, and $\mL$ runs over the host lattice
vectors in $d$ periodic dimensions, with $0\le d\le3$. For $d=0$ the only lattice
vector is $\mL=\mathbf{0}$, and eq~\ref{eq:rienv} is the ordinary molecular fit. The periodic host calculation
determines the coefficients by minimizing the Coulomb norm of the fitting residual. The
charged and chargeless decomposition of
ref~\citenum{Lazarski.Burow.ea:Density.2015} constrains the integral of $\rEnvF$ to the
electron count of the host cell. The embedded guest calculation reads only the coefficients, the
auxiliary basis, the host geometry, and the lattice.
It does not require anything else like molecular orbitals, density matrix, and wave function of the host.
Most importantly, no plane-wave grid, pseudopotential, or periodic calculation of the combined system is required. 

Since the host density is simply a coefficient vector, the
embedded calculation costs about as much as a gas-phase treatment of the molecular active subsystem, and
the source of that vector is flexible. 
Any electronic structure method that
delivers a density fitted in atom-centered Gaussian functions can be used to get the host density. 

The use of fitted density reduces the intersubsystem electrostatic Coulomb integrals to
three-center form and keeps the embedding potential consistent with the host calculation.
The nonadditive exchange-correlation and kinetic terms evaluate the very same $\rEnvF$
pointwise on the quadrature grid, as detailed in Section~\ref{sec:nadd-periodic}. Each nonadditive energy is a
difference of three functional values, and when all three see the host through the same
density, the representation error largely cancels inside the difference instead of
accumulating across mixed representations. The Coulomb-metric fit is
variational for the host Coulomb energy. A cross interaction with a separately fitted
environment does not, by itself, have a general quadratic error bound.

The complete electrostatic part of the embedding potential matrix is given by Coulomb
lattice sums over the combined electronic and nuclear charge distribution of the host,
\begin{equation}
  V^{\mathrm{elstat}}_{\mu\nu}
  = \left(\mu\nu\,\middle|\,\rEnvF - \rho_{\mathrm{n}}^{\mathrm{env}}\right).
  \label{eq:velstat}
\end{equation}
Here $\mu\nu$ is a product of active-subsystem basis functions and
$\rho_{\mathrm{n}}^{\mathrm{env}} = \sum_{\mL}\rho_{\mathrm{n}\mL}$ is the positive nuclear charge density of the host,
with $\rho_{\mathrm{n}\mL}$ that of cell $\mL$. Following ref~\citenum{Lazarski.Burow.ea:Density.2015}, the right-hand
entry is evaluated as a direct-space lattice sum. Equation~\ref{eq:velstat} is the matrix
form of the electrostatic term in eq~\ref{eq:vemb}, with $\rEnv$ replaced by $\rEnvF$.
For $d=0$ the sum reduces to the single term $\mL=\mathbf{0}$, the near field defined
below is the whole finite host, and the far field is empty. Molecular and periodic
environments are therefore the $d=0$ and $d\ge1$ cases of one construction, evaluated by
the same machinery. For the neutral
host cells considered here, every distribution
$\tilde{\rho}_{\mL}-\rho_{\mathrm{n}\mL}$ has zero net charge.

The lattice sums are partitioned, exactly as the Coulomb lattice sums of the periodic
ground-state method,\cite{Lazarski.Burow.ea:Density.2015} into crystal near-field (CNF)
and crystal far-field (CFF) contributions,
\begin{equation}
  \left(\mu\nu\,\middle|\,\rEnvF - \rho_{\mathrm{n}}^{\mathrm{env}}\right)
  = \sum_{\mL\in\mathrm{CNF}}
    \left(\mu\nu\,\middle|\,\tilde{\rho}_{\mL} - \rho_{\mathrm{n}\mL}\right)
  + \sum_{\mL\in\mathrm{CFF}}
    \left(\mu\nu\,\middle|\,\tilde{\rho}_{\mL} - \rho_{\mathrm{n}\mL}\right).
  \label{eq:cnfcff}
\end{equation}
The CNF contains host cells with lattice indices $|n_i|\le k$, where
$i=1,\ldots,d$. Their auxiliary functions and nuclei are replicated explicitly in real
space, shell tables and fitting coefficients included, so the near field is a genuine
finite structure. Its interaction with the active
subsystem is evaluated with the same direct
integration and octree-based CFMM machinery as a molecular
environment.\cite{Burow.Sierka:Linear.2011,Sharma.Sierka:Efficient.2022} The extent $k$
is chosen automatically. Every removed host site and every cell the active subsystem reaches must
lie inside the explicit region, and the code enlarges $k$ until no charge distribution
extends beyond it into cells the far-field operator carries, aborting with a clear
message when the requested geometry cannot satisfy this condition. Before the near field
is built, the host is shifted rigidly by a lattice vector that brings the region of
interest to the home cell. A perfect lattice is invariant under this shift. The
implementation derives it from the removed sites when a void is present and from the
centroid of the guest otherwise, and a guest placed several cells away from the
origin reproduces the home-cell result to all printed digits. During a structure
optimization the shift, the explicit extent, and the site selection of the first step are
frozen and reused (optimization loop in Figure~\ref{fig:scheme}b), because these discrete choices are step functions of the geometry and
re-deriving them at every displaced structure would hand the optimizer a discontinuous
surface. The frozen topology is stored in a small restart file.
Figure~\ref{fig:scheme}a summarizes the construction.

The infinite CFF part is condensed into a single lattice operator. In analogy to the CFF
contribution of the periodic Coulomb
construction,\cite{Lazarski.Burow.ea:Density.2015,Irmler.Burow.ea:Robust.2018} it is
evaluated as the contraction
\begin{equation}
  \sum_{\mL\in\mathrm{CFF}}
  \left(\mu\nu\,\middle|\,\tilde{\rho}_{\mL} - \rho_{\mathrm{n}\mL}\right)
  = \omega^{\mu\nu} \otimes \mu^{\tilde{\rho}-\rho_{\mathrm{n}}}
  = \sum_{l=0}^{L_{\max}} \sum_{m=-l}^{l}
    \omega^{\mu\nu}_{lm}\, \mu^{\tilde{\rho}-\rho_{\mathrm{n}}}_{lm},
  \label{eq:cff}
\end{equation}
which contracts the multipole moments $\omega^{\mu\nu}_{lm}$ of the product $\mu\nu$
with the Taylor moments $\mu^{\tilde{\rho}-\rho_{\mathrm{n}}}_{lm}$ of the CFF lattice
array. The truncation order $L_{\max}$ is that of the host CFMM calculation. Both sets of
moments use the same center in the active subsystem. The Taylor
moments are generated from the neutral host-cell distribution through external-to-local
translation operators and their lattice sums.\cite{Lazarski.Burow.ea:Density.2015} The
same convergence treatment is used as in the periodic Coulomb implementation. In three
dimensions, cells with a nonzero dipole moment require the established dipole
correction.\cite{Lazarski.Burow.ea:Density.2015,Irmler.Burow.ea:Robust.2018}

The current test energies are invariant within $10^{-8}$~hartree when host atoms are reassigned
to equivalent home cells. Replacing a primitive cell by a supercell changes the embedding
energy by values between $3\times10^{-5}$ and $9\times10^{-5}$~hartree for the test
hosts. This comparison measures the combined sensitivity to the fitted representation and
the numerical setup. It is not a formal estimate of the fitting error alone. Analogous
CNF and CFF contractions with $\rho_{\mathrm{n}}^{\mathrm{act}}$ in place of $\mu\nu$
give the attractive interaction
$-(\rho_{\mathrm{n}}^{\mathrm{act}}\,|\,\rEnvF)$ and the repulsive interaction
$(\rho_{\mathrm{n}}^{\mathrm{act}}\,|\,\rho_{\mathrm{n}}^{\mathrm{env}})$ in
eq~\ref{eq:eint}. The energy, KS matrix, and gradient use the same CNF lists and CFF
operator. This shared construction makes the three expressions algebraically consistent.

Increasing the explicit region reduces the multipole truncation error in the CFF\@. For the
current test systems, embedding energies from successive values of $k$ agree within
$10^{-9}$~hartree once the automatically selected value is reached.

\subsubsection{Nonadditive terms for a periodic environment}
\label{sec:nadd-periodic}

The nonadditive matrices $\mathbf{X}_{\mathrm{nadd}}$ and $\mathbf{T}_{\mathrm{nadd}}$
of eq~\ref{eq:vembmat}, and the corresponding energies in eqs~\ref{eq:naddxc}
and~\ref{eq:naddke}, require the pointwise host density on the quadrature grid rather
than its electrostatic potential. The
exchange-correlation and kinetic functionals used here are semilocal, so the integrands
of the differences in eqs~\ref{eq:naddxc} and~\ref{eq:naddke} vanish wherever
$\rAct$ and $\rEnv$ do not overlap. Gaussian densities decay exponentially with
distance. Only host atoms in the immediate vicinity of the active subsystem therefore
contribute, and no lattice summation is required for the nonadditive terms.

The quadrature spans the active atoms together with all host atoms, periodic images
included, that lie within a user radius $R$ of any active atom (green region in Figure~\ref{fig:scheme}a). Removed sites are
excluded from this selection. The Stratmann-Scuseria-Frisch partition weights\cite{Stratmann.Scuseria.ea:Achieving.1996}
are renormalized over exactly this union of centers, so the quadrature remains an exact partition of unity and
its accuracy is limited by the grid size alone. Interior weights coincide with the
crystalline ones, while the outermost selected atoms absorb the exterior. This boundary
contribution belongs to the $R$ truncation error, which decays with the overlap of the
densities. In practice the total energies of the applications below change by well under
$10^{-4}$~hartree between $R=0$, where only the active grids are kept, and $R=10$~bohr.

The host density and its gradient on the union grid come from the fitted density alone.
One tree traversal before the SCF evaluates the void-corrected $\rEnvF$ of
eq~\ref{eq:deficit} with its Cartesian gradient at every point and caches the values in
the grid ordering. For gradient calculations the same pass also stores the six second
derivatives. The cost of this step is small, a few percent of one SCF iteration, and it
is paid once. During the SCF the frozen values are simply added to the freshly computed
active density in each point batch.

The nonadditive quantities are evaluated in a single pass over that grid. At every point
batch the active density is formed from its density matrix, the cached host values are
added pointwise, and the functional is evaluated for $\rAct+\rEnvF$, $\rAct$, and $\rEnvF$
in immediate succession, so that the energy density and the potential entering
$\mathbf{X}_{\mathrm{nadd}}$ and $\mathbf{T}_{\mathrm{nadd}}$ are assembled as
differences inside the integrand, not as differences of separately integrated terms.
Numerically, the difference integrand is screened out
wherever $\rAct$ is negligible, so quadrature regions that contribute
nothing are skipped as whole octree branches, and no cancellation between large
separate integrals occurs. For the derivatives, the quadrature-weight response of the
reference $E[\rEnvF]$, which in a separate-integral formulation must be evaluated and
subtracted explicitly, cancels inside the integrand by construction.

Functionals of the meta generalized gradient approximation (meta-GGA) need one further
ingredient, the kinetic energy density
$\tau=\tfrac12\sum_i|\nabla\psi_i|^2$. For the active subsystem it is exact, since the
orbitals are present. The host contributes no orbitals, only its fitted density, so
$\tau^{\mathrm{env}}$ is taken from an orbital-free model evaluated pointwise from the
cached $\rEnvF$ and $\nabla\rEnvF$. All model expressions are built from one smooth
regularization of the fitted density,
\begin{equation}
  g(\rho) = \tfrac12\left(\rho + \sqrt{\rho^2 + \delta_\rho^2}\,\right),
  \label{eq:taureg}
\end{equation}
which tends to $\rho$ for $\rho\gg\delta_\rho$ and to zero for negative $\rho$, with
$\delta_\rho=10^{-10}$. With $s=|\nabla\rho|^2$, the Thomas-Fermi and von Weizs\"acker
densities read\cite{Thomas:Calculation.1927,Fermi:Statistische.1928,Weizsaecker:Zur.1935}
\begin{equation}
  \tau_{\mathrm{TF}} = C_{\mathrm{TF}}\, g^{5/3},
  \qquad
  \tau_{\mathrm{W}} = \frac{s\,g}{8\left(g^2+\delta_{\mathrm{W}}^2\right)},
  \label{eq:tautfvw}
\end{equation}
with $C_{\mathrm{TF}}=\tfrac{3}{10}(3\pi^2)^{2/3}$ and
$\delta_{\mathrm{W}}=10^{-8}$. A
fitted density is a least-squares object and undershoots into small negative pockets in
the intermolecular tails, of the order of $10^{-3}$ in our tests on molecular crystals.
There the bare quotient $s/(8\rho)$ diverges while the density itself carries no weight.
The damped form of eq~\ref{eq:tautfvw} agrees with the bare one to a relative error of
order $(\delta_{\mathrm{W}}/g)^2$ at every physical density and falls to zero together with
the density, which is the correct limit. The default host model is the second-order
gradient expansion\cite{Brack.Jennings.ea:On.1976} made Pauli safe,
\begin{equation}
  \tau^{\mathrm{env}}
  = \tau_{\mathrm{W}}
  + h\!\left(\tau_{\mathrm{TF}} - \tfrac{8}{9}\,\tau_{\mathrm{W}}\right),
  \qquad
  h(p) = \tfrac12\left(p + \sqrt{p^2+\delta_\tau^2}\,\right),
  \label{eq:taugea}
\end{equation}
with $\delta_\tau=10^{-12}$, which guarantees
$\tau^{\mathrm{env}}\ge\tau_{\mathrm{W}}$ everywhere, a bound that
functionals of the TPSS family assume,\cite{Tao.Perdew.ea:Climbing.2003} and reduces to
the plain gradient expansion wherever that expansion is valid. Thomas-Fermi and von
Weizs\"acker alone are available as alternatives. These statements hold for densities well above the
regularization scales $\delta_\rho$, $\delta_{\mathrm{W}}$ and $\delta_\tau$, which is the case wherever the
density carries weight. The analytic derivatives of the
implemented model expressions, chain rule through eqs~\ref{eq:taureg}
to~\ref{eq:taugea} included, feed the gradient, so energy, potential, and forces stay
mutually consistent. The kinetic energy functional, of generalized gradient
approximation (GGA) class throughout, never sees the $\tau$ variable and runs on a
packed view of the same data.

The CNF extent is
selected by the long-ranged electrostatic criterion of Section~\ref{sec:periodic},
whereas the density overlap is exponentially short ranged. A CNF that satisfies the CFF
convergence radius is therefore also sufficient for the nonadditive terms, and the
$k$-convergence of the total embedding energies quoted above covers the electrostatic
and nonadditive contributions alike.

\subsubsection{Void construction in the host}
\label{sec:defect}
A vacancy, a substitution, or an active subsystem that includes host atoms changes only a local part of the host, yet periodic
calculations of defective crystalline hosts normally require a supercell. 
The present scheme's advantage is that it removes the supercell requirement, and it does so uniformly for every
dimensionality. 
The user lists the coordinates of
the host atoms to be removed (void branch in Figure~\ref{fig:scheme}b). Each entry is matched against the host structure, with a tight
tolerance and an element check, and resolved to a pair of cell atom and periodic image.
For a molecular host the image is trivial and the removal simply deletes atoms. 
For a chain, a slab, or a crystal the removed set may span several
cells. A molecule can therefore
replace its complete motif in a molecular crystal even when that motif straddles a cell
boundary.

Let $\mathcal{V}$ denote the resolved sites. The environment density seen by the active subsystem is
\begin{equation}
  \rEnvF_{\mathcal{V}}(\mr)
  = \sum_{(\alpha,\mL)\notin\mathcal{V}}
    c_\alpha\,\alpha_{\mL}(\mr),
  \label{eq:deficit}
\end{equation}
and the nuclear charges of the same sites are removed from the source distribution. The
fit is partitioned over atoms by construction, so eq~\ref{eq:deficit} deletes exactly the
atom-centered contributions of the removed sites and nothing else. Every removal is
finite and lies inside the explicit CNF, whose replicated coefficient vector simply
carries zeros on the removed slices. The CFF retains the periodicity of the perfect host
and its per-cell neutrality, which the renormalized lattice operator requires. 
The embedded calculation checks this neutrality in two ways: against the electron count
recorded by the host calculation and by direct integration of the fit coefficients, and
stops if either check fails.

Equation~\ref{eq:deficit} is an algebraic partition of the fitted perfect-host density,
not a relaxed density of the defective crystal. The atom-centered terms are not
separately observable densities, and their sum is not guaranteed to be nonnegative
pointwise. The implementation therefore prints the minimum of the fitted density and its
integrated negative part on the quadrature grid, together with the removed charge per
site. The deleted terms are
whole atomic contributions of a Coulomb-metric fit, so the void carries the full charge
of the removed atoms exactly. Removing an entire motif and
reinserting the same motif as the active subsystem must reproduce the corresponding
molecule-in-perfect-crystal result, and both nonadditive energies must vanish. In our
tests they vanish to all printed digits, and the removed electron count is reproduced
exactly. 
What the frozen partition leaves out is the relaxation of the host density around the
void. 
This approximation
is common to all frozen-density schemes and could be lifted only by updating the host
density in response to the active subsystem.
Its effect on the guest
diminishes as more host atoms are promoted to the active subsystem, as the comparison
with periodic calculations of the combined systems for CO and NH$_3$ on MgO(001) shows
(Sections~\ref{sec:co-mgo} and~\ref{sec:applications}).

The complete embedding functional, the nonadditive terms of
Section~\ref{sec:nadd-periodic} included, applies to the void geometry unchanged, and so
does the analytic gradient of Section~\ref{sec:gradients}.

\subsection{Analytic nuclear gradients}
\label{sec:gradients}

For a frozen $\rEnv$, the energy in eq~\ref{eq:etot} depends explicitly on the active
nuclear positions $\mR_a$ through the integrals and internuclear terms. It also depends
implicitly on them through the converged SCF density. At SCF stationarity, the latter
dependence gives the usual overlap contribution,
\begin{equation}
  \frac{\mathrm{d}E^{\mathrm{tot}}}{\mathrm{d}\mR_a}
  = \left.\frac{\partial E^{\mathrm{tot}}}{\partial \mR_a}\right|_{\matD}
  - \Tr\!\left[\matW\, \frac{\partial \matS}{\partial \mR_a}\right],
  \label{eq:gradtotal}
\end{equation}
where $\matD$ and $\matS$ are the active-subsystem density and overlap matrices.
$\matW$ is the energy-weighted density matrix formed with eigenvalues of the embedded KS
matrix, which includes $\matV_{\mathrm{emb}}$ from eq~\ref{eq:vembmat}. The fixed-density
partial derivative contains the standard KS-DFT terms of the isolated active
subsystem\cite{Lazarski.Burow.ea:Density.2016} and derivatives of every interaction term
in eq~\ref{eq:eint}.

\paragraph{Electrostatic interaction with the environment density and nuclei.}
The derivative of
$\sum_{\mu\nu}D_{\mu\nu}
(\mu\nu\,|\,\rEnvF-\rho_{\mathrm{n}}^{\mathrm{env}})$ contains derivative Coulomb
integrals of $\mu\nu$ against the frozen host sources. The CNF contribution is evaluated
with the derivative engine of the DF-CFMM Coulomb
gradient.\cite{Lazarski.Burow.ea:Density.2016} That engine differentiates a complete
Coulomb energy expression. The implementation forms a difference of derivative passes,
with the active density matrix present in one and zero in the other. Contributions already
included in the isolated active-subsystem gradient are excluded. The remaining terms are
the cross derivatives between $\rAct$ and $\rEnvF$ and between $\rAct$ and
$\rho_{\mathrm{n}}^{\mathrm{env}}$. The CFF contribution is the derivative form of
eq~\ref{eq:cff}. It contracts the Taylor moments with derivative multipole moments
$\omega^{(\mu\nu)'}_{lm}$, as in the periodic Coulomb
gradient.\cite{Lazarski.Burow.ea:Density.2016}

\paragraph{Active nuclei in the environment density.}
The attractive term $-(\rho_{\mathrm{n}}^{\mathrm{act}}\,|\,\rEnvF)$ is independent of
$\matD$. Its gradient is assembled from derivative nuclear-attraction integrals over the
host auxiliary functions in the CNF\@. For a periodic environment, it also contains the
derivative of the CFF point-charge contraction translated to each active nucleus.

\paragraph{Internuclear repulsion.}
The term $(\rho_{\mathrm{n}}^{\mathrm{act}}\,|\,\rho_{\mathrm{n}}^{\mathrm{env}})$
gives explicit Coulomb derivatives from the CNF images. In a periodic environment, it
also includes the derivative of the CFF point-charge contraction. The energy and gradient
use the same lattice operator.

\paragraph{Nonadditive exchange-correlation and kinetic terms.}
In a monomolecular basis, $E[\rEnv]$ is independent of the active nuclear positions. For
a molecular environment, the derivative of each nonadditive term is the difference of
two standard grid-based derivatives. One is evaluated for the total density on the
total-system grid and retained only on active nuclei. The other is evaluated for the
active density on the active-subsystem grid. The kinetic contribution uses the same
machinery with a KEDF in place of the exchange-correlation functional.

For a crystalline environment the nonadditive terms live on the union grid of
Section~\ref{sec:nadd-periodic}, and their derivative decomposes into four
contributions. The first is the basis term, the derivatives of the active basis
functions contracted against the summed difference potential. The second corresponds to the
motion of the grid representation of $\rAct$ itself, which is exact by translational
invariance of a purely atom-centered density. The third has no analogue in a standard
DFT gradient. Quadrature points belonging to the active atoms move rigidly with them, while
$\rEnvF$ stays fixed in space, so the frozen density changes along the path of every
moving point,
\begin{equation}
  \frac{\mathrm{d}\rEnvF(\mr_p)}{\mathrm{d}\mR_a}
  = \nabla\rEnvF(\mr_p)
  \qquad \text{for points } p \text{ owned by atom } a.
  \label{eq:nodemotion}
\end{equation}
The corresponding gradient contribution contracts the difference of the total and the
environment functional derivatives with this frozen-field motion,
\begin{equation}
  \sum_{p\in a} w_p\,
  \Bigl\{
  \bigl[v(\rTot)-v(\rEnvF)\bigr]_p \nabla\rEnvF(\mr_p)
  + \bigl[\bm{v}_\sigma(\rTot)-\bm{v}_\sigma(\rEnvF)\bigr]_p\,
    \mathbf{H}[\rEnvF](\mr_p)
  \Bigr\},
  \label{eq:nodeterm}
\end{equation}
where $w_p$ is the quadrature weight of point $p$, $v$ collects the density
derivatives of the exchange-correlation and kinetic
functionals, $\bm{v}_\sigma$ their derivatives with respect to the density gradient $\nabla\rho$, and $\mathbf{H}[\rEnvF]$ is the
cached Hessian of the fitted density. For meta-GGA functionals the analogous $\tau$ term
appears, with $\nabla\tau^{\mathrm{env}}$ assembled from the exact derivatives of the
model of eqs~\ref{eq:taureg} to~\ref{eq:taugea} and the same cached Hessian. The
fourth contribution consists of the quadrature weight derivatives of the union
partition, which multiply the full pointwise nonadditive energy density. The weight
derivatives of the environment centers are discarded, since those centers do not move.
The implementation enables the weight derivatives automatically whenever the gradient of
an embedded subsystem is requested.

In a supersystem basis, some environment orbitals can use basis functions centered on
active atoms. The energy of the environment then has explicit basis-center derivatives.
These terms cannot be dropped from eqs~\ref{eq:naddxc} and~\ref{eq:naddke}. The
implementation evaluates and subtracts the environment contributions, including the
corresponding correction to the auxiliary Coulomb derivative pass.

\paragraph{Rigid-body components of the gradient.}
For an isolated molecule, translational and rotational invariance makes the net force
and torque vanish, and these components are often projected out of the gradient. The
frozen host breaks this invariance: the position and orientation of the active
subsystem relative to the host change the embedded energy, so the net force and torque
are physical. They are therefore retained in the embedded gradient, and the optimizer
must not remove the rigid translations and rotations of the active subsystem. The individual gradient terms were checked against frozen-density finite differences,
in which the density matrix and fitting coefficients are held fixed while the
integrals, grids, and lattice operators are rebuilt at each displaced geometry. The
deviations are at most $10^{-7}$~$E_{\mathrm{h}}\,a_0^{-1}$ for every term. The complete
gradient is tested by the structure relaxations of Section~\ref{sec:applications}.



The assembly covers any explicit CNF extent $k$. The near-field derivative passes run
over the replicated host cells. The CFF contribution differentiates the same
cell-excluding lattice operator as the energy. The attraction between the active nuclei
and the fitted host density of each explicit cell is differentiated by a dedicated
kernel that evaluates the vertical recurrences of the corresponding energy integrals one
order higher. Each contribution therefore differentiates exactly the lattice sum that
the energy evaluates. A molecular active subsystem embedded in a primitive host cell has analytic
gradients without any supercell construction.

\subsection{Extensions to WFT-in-DFT and RT-TDDFET}
\label{sec:extensions}

The embedding potential of eq~\ref{eq:vembmat} is a one-electron operator over the basis functions of the active subsystem. After convergence of the embedded DFT calculation, its matrix is exported to a file (Figure~\ref{fig:scheme}b). 
For WFT-in-DFT, the exported matrix is added to the Hartree-Fock core Hamiltonian of the active subsystem, and the correlated calculation proceeds on the resulting embedded reference orbitals, as in ref~\citenum{Sharma.Sierka:Efficient.2022}. 
The potential is computed once from the DFT densities of both subsystems and kept fixed during the Hartree-Fock and the correlated calculation. 
Any post-Hartree-Fock model available in the molecular modules can therefore be applied without modification, since the environment enters only through the fixed one-electron matrix. 
In the present work this route is exercised with CC2 excitation energies,\cite{Christiansen.Koch.ea:second-order.1995,Hattig.Weigend:CC2.2000} as in our earlier CC2-in-DFT studies of solvated molecules and ionic materials,\cite{Sharma.Sierka:Efficient.2022,Sharma.Sierka:Optical.2024} but now utilizing the direct-space potential of Section~\ref{sec:periodic}. 
The excitation energies follow from the linear-response CC2 equations of the embedded active subsystem, and the ground-state potential is used for the excited states as well. 
The environment density therefore does not respond to the active subsystem's excitation. 
This state-independent approximation, which neglects the differential polarization of the environment, is appropriate for excitations localized on the active subsystem.\cite{Daday.König.ea:State-Specific.2013,Sharma.Sierka:Efficient.2022}
The construction of the embedded reference is identical for molecular and periodic environments, including hosts with voids. 
No periodic calculation of the combined system is required at any stage.
Analytic WFT-in-DFT gradients are not yet available: the missing terms are the contractions of the derivative machinery of Section~\ref{sec:gradients} with the relaxed correlated density instead of the SCF density.

For real-time propagation, the time-dependent KS equations of the active subsystem are integrated with the embedding potential included in the KS matrix, following our previous RT-TDDFET scheme~\cite{Mueller.Sharma.ea:Real-time.2020, Li.Saleh.ea:Resonance.2023,Sharma.Sierka:Efficient.2022}. 
The embedding potential is held fixed at its ground-state form during the propagation, which corresponds to the state-independent approximation. 
The direct-space potential makes this propagation available for molecular chromophores in periodic environments without any periodic calculation of the combined system. 
Section~\ref{sec:rt-results} compares this construction with the earlier embedding-potential variant of ref~\citenum{Sharma.Sierka:Efficient.2022} against a linear-response TDDFT reference for the combined system.

\subsection{Implementation notes}
\label{sec:implementation}

The scheme is implemented in the \textsc{riper} module of
TURBOMOLE.\cite{Franzke.Holzer.ea:TURBOMOLE.2023,Lazarski.Burow.ea:Density.2015} The
workflow has two stages (Figure~\ref{fig:scheme}b). The host calculation, run once, exports the fitting
coefficients and the energies after convergence. 
The embedded
calculation reads them, resolves the removed sites (if any), builds the near field and the
far-field operator, caches the frozen density on the union grid, and then runs an
ordinary molecular SCF of the active subsystem in which the precomputed electrostatic
matrix ($\matV^{\mathrm{elstat}}$ of eq~\ref{eq:velstat}, including the CFF operator)
is added to every Fock matrix and the nonadditive potential is rebuilt each iteration.
The Coulomb term of the active subsystem itself uses the multipole accelerated
resolution of the identity Coulomb engine (MARI-$J$),\cite{Sierka.Hogekamp.ea:Fast.2003} exactly as a gas-phase molecular
calculation would. The embedded gradient adds
derivative Coulomb passes over the replicated near field, far-field derivative
contractions, and the four grid-based contributions of Section~\ref{sec:gradients}.
These operations reuse the OpenMP-parallel CFMM and quadrature code.\cite{Burow.Sierka:Linear.2011,Lazarski.Burow.ea:Density.2015,%
Lazarski.Burow.ea:Density.2016}

The molecular character of the embedded SCF brings the full feature set of the
molecular code along at no extra cost. Spin-polarized active subsystems, global and
range-separated hybrid functionals, whose exact exchange runs over the molecular active subsystem alone
while the nonadditive terms stay semilocal, dispersion corrections, fractional
occupations with Gaussian smearing, and effective core potentials all work unchanged.
Hosts whose atoms carry effective core potentials are supported as well. Their valence
fit and effective charges act electrostatically, and the semilocal pseudopotential
operator of every explicit-zone atom joins the embedding matrix together with its
gradient, so the active density feels the short-range repulsion of the frozen cores.
Meta-GGA functionals are available on both sides of the partition through the host
$\tau$ model of Section~\ref{sec:nadd-periodic}. The host calculation may employ any
$k$-point mesh. The embedded calculation inherits only the fitted density, whose quality
the mesh improves.

Cost is dominated by the one-time host calculation. Everything the embedding adds to
the molecular SCF, the static electrostatic matrix, the cached density pass, and the
per-cycle nonadditive quadrature, scales with the size of the active subsystem and its
immediate surroundings, not with the host. Over a meta-GGA host the nonadditive quadrature evaluates the costlier
kernels at three densities together with the host $\tau$ model, which roughly triples
that step and is only visible for small active subsystems. The removed sites are
specified as a coordinate block in the input, the selection radius $R$ and the
near-field extent have working defaults, and the converged embedding potential is
written to a file for the WFT-in-DFT and RT-TDDFET extensions of
Section~\ref{sec:extensions}. The \texttt{riperembed} driver automates the two
stages.\cite{PhD-Sharma:2024}

\subsection{Interaction energy of CO and MgO(001) using embedding: accuracy
and convergence with cluster size}
\label{sec:co-mgo}
To investigate the accuracy of our embedding method and validate the implementation, we examine the adsorption energetics of the CO molecule on the MgO(001) surface, a standard weak-adsorption benchmark for surface theory.\cite{Ye.Berkelbach:Adsorption.2024} 
The interaction energy predicted by PBE (without dispersion correction) is known to be $\sim$16~kJ/mol.\cite{Bucko.Lebegue.ea:Extending.2014, Pasti.Skorodumova:Structural.2016}
So if our embedding scheme aims to reproduce this value through PBE-in-PBE (molecule-in-periodic embedding), it must exhibit errors an order of magnitude smaller than that.

The structure consists of the CO molecule being C-down on a five-fold coordinated surface Mg. 
The substrate contains 64 atoms.
The clean slab is treated as the host for embedding.
The active subsystem is made by carving a cluster that includes the CO molecule (guest) and the nearby MgO atoms of the host and uses the monomolecular basis. 

Since the substrate atoms are included in the active region, the void construction scheme of Section~\ref{sec:defect} is employed (void branch of Figure~\ref{fig:scheme}b). 
To investigate the effect of cluster size, we consider various active subsystems with an increasing number of host MgO atoms included in the active subsystem, specifically CO$+$Mg$_5$O$_5$, CO$+$Mg$_7$O$_7$, and CO$+$Mg$_9$O$_9$.
All three span both layers of the slab.
Each promoted host atom is removed from the fitted host density by eq~\ref{eq:deficit} and reappears with its own basis functions as a part of the active subsystem.

\begin{figure}
  \centering
  \includegraphics[width=0.5\linewidth]{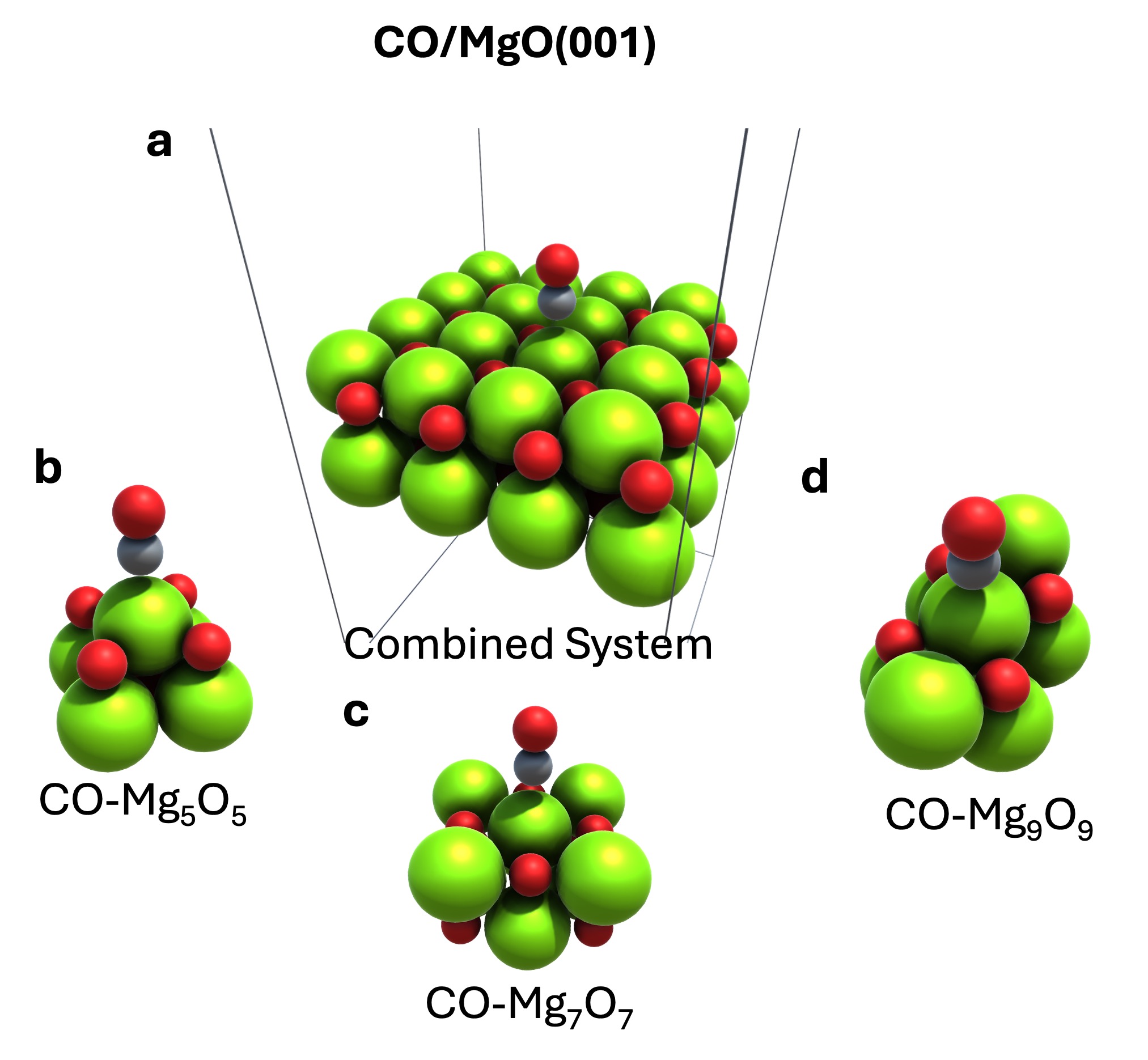}
  \caption{CO on MgO(001). \textbf{a}, The combined system of 66 atoms in the two-dimensionally
  periodic surface cell. \textbf{b}--\textbf{d}, The three active subsystems
  CO$+$Mg$_5$O$_5$ (\textbf{b}), CO$+$Mg$_7$O$_7$ (\textbf{c}) and CO$+$Mg$_9$O$_9$ (\textbf{d}).}
  \label{fig:co-mgo-system}
\end{figure}

Figure~\ref{fig:co-mgo-system} shows the combined system and the three active subsystems considered for embedding.
The interaction energy is scanned rigidly. 
CO is held at
$r(\mathrm{C}\text{-}\mathrm{O}) = 1.14$~\AA\ and translated along the surface normal
above the site Mg with every substrate atom fixed, over fifteen separations from 1.90 to
6.00~\AA. 

For the embedded curves the interaction energy at separation $d$ is
\begin{equation}
  \Eint_{\mathrm{emb}}(d)
  = E^{\mathrm{tot}}\!\left[\rAct_{\mathrm{CO}+\mathcal{C}}(d),\,
      \rEnvF_{\mathcal{V}}\right]
  - E^{\mathrm{tot}}\!\left[\rAct_{\mathcal{C}},\,\rEnvF_{\mathcal{V}}\right]
  - E\!\left[\rho^{\mathrm{CO}}\right],
  \label{eq:eint-emb}
\end{equation}
where $\mathcal{C}$ is the promoted cluster, $\mathcal{V}$ the corresponding void of
eq~\ref{eq:deficit}, and $E[\rho^{\mathrm{CO}}]$ the gas-phase CO energy at the same
internal geometry. 
The reference curve is the ordinary periodic expression for the
combined system, the supersystem,
\begin{equation}
  \Eint_{\mathrm{sup}}(d)
  = E^{\mathrm{per}}\!\left[\mathrm{CO}/\mathrm{slab}\right](d)
  - E^{\mathrm{per}}\!\left[\mathrm{slab}\right]
  - E\!\left[\rho^{\mathrm{CO}}\right],
  \label{eq:eint-sup}
\end{equation}
where the first two terms are obtained from periodic DFT calculations.

Both combined (reference) and embedded interaction energies are counterpoise (CP) corrected for basis set superposition error in the Boys-Bernardi
sense.\cite{Boys.Bernardi:calculation.1970} The correction has a simple structure here.
Because the frozen host enters only through its fitted density and carries no basis
functions, the sole superposition channel is the adsorbate against the promoted cluster,
and it lies entirely inside the active subsystem. Writing $\mathcal{S}$ for the substrate
fragment of a given case, either the promoted cluster or the periodic slab,
\begin{equation}
  \Eint_{\mathrm{CP}}(d)
  = E\!\left[\mathrm{CO}+\mathcal{S}\right](d)
  - E\!\left[\mathcal{S}\right]^{\mathrm{CO\ ghost}}(d)
  - E\!\left[\mathrm{CO}\right]^{\mathcal{S}\ \mathrm{ghost}}(d),
  \label{eq:cp}
\end{equation}
with both fragment terms evaluated in the supersystem basis of the complex at that separation. 
At the minimum the counterpoise correction is 3.9~kJ/mol for the largest 20-atom active subsystem against 3.3~kJ/mol for the 66-atom combined system, so the
embedded cluster gains more from the adsorbate basis functions than the full slab does.

\begin{figure}
  \centering
  \includegraphics[width=0.8\linewidth]{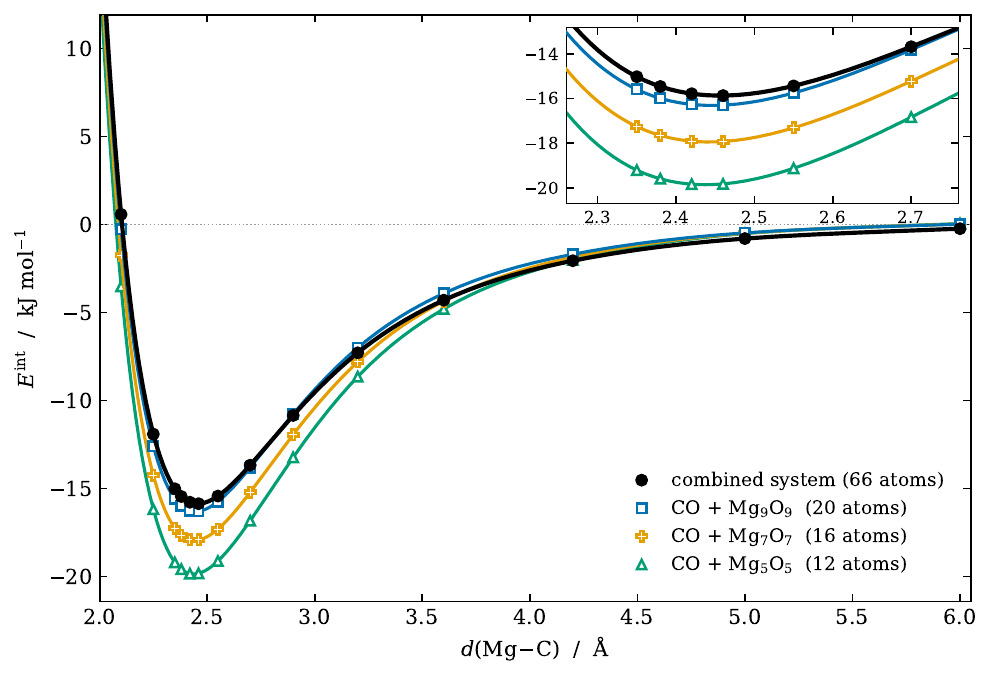}
  \caption{Counterpoise-corrected interaction energy of CO on MgO(001) along the surface
  normal, PBE/def2-TZVP without dispersion, for the three active subsystems and for the
  periodic calculation of the combined system. Markers are the 15 computed separations,
  and lines are splines through them. The inset magnifies the well. The embedded curves
  converge to the reference from below as the promoted region grows, and all curves
  coincide beyond 4~\AA, where the interaction vanishes and the embedding potential can no
  longer affect the result.}
  \label{fig:co-mgo-pes}
\end{figure}

Figure~\ref{fig:co-mgo-pes} shows the resulting curves and Table~\ref{tab:co-mgo}
reports the explicit numbers. Convergence with the cluster size of the active subsystem is fast and monotonic. 
The well depth deviates from the combined-system reference by $-4.00$, $-2.07$, and $-0.44$~kJ/mol
for the DFT-in-DFT calculations using the CO$+$Mg$_5$O$_5$, CO$+$Mg$_7$O$_7$, and CO$+$Mg$_9$O$_9$ clusters as active subsystems respectively.
Similarly, the equilibrium separation differs from the combined-system reference by $-0.022$, $-0.017$,
and $-0.012$~\AA, respectively. 
The embedded DFT calculation with the CO$+$Mg$_9$O$_9$ cluster as the active subsystem therefore reproduces the reference $-15.87$~kJ/mol
interaction energy to 0.44~kJ/mol and the whole curve to 0.50~kJ/mol root mean square, while
treating fewer than a third of the atoms of the combined system quantum mechanically and only at the molecular level.
Both sequences approach the reference from the same side. 
The error shrinks as more MgO atoms are promoted to the active region.

The distance dependence of the error identifies its origin and validates the
implementation. At large separation the interaction vanishes, so the embedding potential
must not affect the result, whatever cluster carries it. The data confirm this. Beyond
3.20~\AA\ the deviation of every curve falls below 1.4~kJ/mol, and beyond 4.20~\AA\ all
three collapse onto a common residual of 0.26 to 0.37~kJ/mol that no longer depends on the
size of the promoted region. 

Furthermore, our counterpoise-corrected combined-system value, $-15.87$~kJ/mol at
2.460~\AA, agrees with the plane-wave PBE results of Bu\v{c}ko et
al.\cite{Bucko.Lebegue.ea:Extending.2014} ($-16.4$~kJ/mol at 2.45~\AA) and of Pa\v{s}ti and
Skorodumova\cite{Pasti.Skorodumova:Structural.2016} ($-16.5$~kJ/mol at 2.43~\AA) to better than 0.7~kJ/mol in energy
and 0.03~\AA\ in equilibrium separation. 
Interestingly, the embedded results ($E^\mathrm{int}=-16.31$ at 2.448~\AA) from the largest active subsystem are in even closer agreement ($<0.1$~kJ/mol) with the reference plane-wave results. 

\begin{figure}
  \centering
  \includegraphics[width=0.7\linewidth]{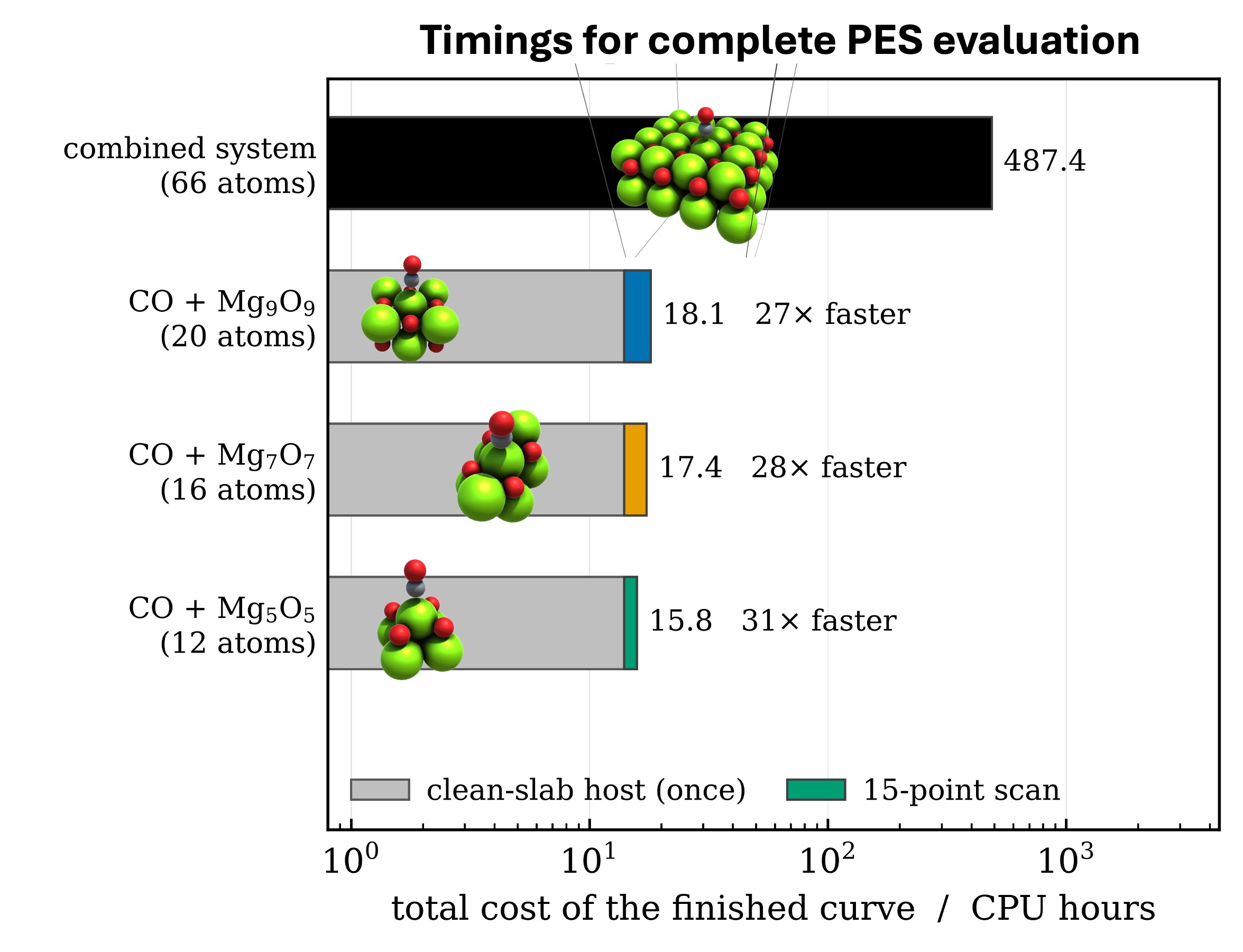}
  \caption{Computational cost of the 15-point interaction-energy curve of CO on MgO(001) in CPU hours (PBE/def2-TZVP, logarithmic scale). The embedded bars stack the one-time clean-slab host calculation (grey) and the 15 embedded points (coloured). The combined-system bar is the 15 periodic calculations.}
  \label{fig:co-mgo-timings}
\end{figure}

The computational cost (CPU hours) of the 15 data points for the interaction energy curve is shown in Figure~\ref{fig:co-mgo-timings}. 
A single periodic calculation of the clean slab produces the fitted density,
which is then utilized by every point of every embedded curve. 
Therefore, no host calculation is repeated for the scan.
Each embedded run contains exactly one SCF, that of the molecular active subsystem in the presence of the embedding potential. 
Generating the 15-point curve for the largest cluster case costs a mere 4.1 CPU
hours on top of the 14.0 CPU hour host, compared to 487 CPU hours needed for evaluating the same 15 points using the combined system.
Embedding results in a speed-up of a factor of 27, including the one-off host calculation. 
Therefore, embedding is especially useful in cases when one is performing a scan, finding favorable adsorption sites, or optimizing the adsorbate geometry. 
Once the host exists, an additional point costs only 16.4 CPU minutes.

\begin{table}
  \centering
  \caption{Counterpoise-corrected interaction energies of CO on MgO(001) from embedded
  active subsystems of growing size, against the periodic calculation of the combined
  system, PBE/def2-TZVP. $d_{\mathrm{min}}$ and $\Eint$ are the position and depth of the
  minimum of the spline through the 15 separations from 1.90 to 6.00~\AA, $\Delta\Eint$ is
  embedded minus combined system, and RMS is the root-mean-square deviation from the
  combined-system curve over all 15 separations.}
  \label{tab:co-mgo}
  \small
  \begin{tabular}{lrrrrrr}
    \toprule
    active subsystem & atoms & basis fn. & $d_{\mathrm{min}}$/\AA & $\Eint$/kJ mol$^{-1}$
    & $\Delta\Eint$/kJ mol$^{-1}$ & RMS/kJ mol$^{-1}$ \\
    \midrule
    CO$+$Mg$_5$O$_5$ & 12 & 377  & 2.438 & $-19.87$ & $-4.00$ & 3.09 \\
    CO$+$Mg$_7$O$_7$ & 16 & 503  & 2.443 & $-17.94$ & $-2.07$ & 1.64 \\
    CO$+$Mg$_9$O$_9$ & 20 & 629  & 2.448 & $-16.31$ & $-0.44$ & 0.50 \\
    \midrule
    combined system  & 66 & 2078 & 2.460 & $-15.87$ & --      & --   \\
    \bottomrule
  \end{tabular}
\end{table}

\subsection{Structure relaxation of the embedded adsorbate: NH$_3$ on MgO(001)}
\label{sec:applications}

In the previous section, we showed that the embedding methodology successfully reproduces the combined-system interaction energies at negligible cost and error. 
We now test the accuracy of the analytic gradients of the embedded active subsystem  (Section~\ref{sec:gradients}) and calculate the adsorption energy by relaxing the adsorbate in the presence of the frozen periodic potential, and compare the results with a regular periodic reference.
We also investigate whether the relaxed structure and
the adsorption energy converge with the cluster size in the same way the rigid scan did.
For this study, we use NH$_3$ adsorption on MgO(001) rather than CO because it is more complex as it contains more degrees of freedom (internal coordinates)
and an orientation.

As in the previous study in Section~\ref{sec:co-mgo}, the MgO slab is treated as the host.
In the initial structure, NH$_3$ sits on MgO with N-down on a five-fold coordinated surface Mg atom and the combined system consists of 68 atoms.
The active subsystem consists of NH$_3$ (guest) and the nearby MgO atoms.
The MgO clusters included in the active subsystem (for cluster-size convergence) are also the same as the previous study.
In every run only the four atoms of NH$_3$ are free and every substrate atom is held fixed at
the reference geometry, in the embedded calculations and in the periodic reference alike.
Therefore the reference and embedded calculations are comparable.
Optimizations use a Cartesian gradient threshold of 0.02~eV/\AA.
The adsorption energy of an embedded run obtained after relaxation of the adsorbate with other atoms fixed is
\begin{equation}
  E_{\mathrm{ads}}^{\mathrm{emb}}
  = E^{\mathrm{tot}}\!\left[\rAct_{\mathrm{NH_3}+\mathcal{C}},\,
      \rEnvF_{\mathcal{V}}\right]
  - E^{\mathrm{tot}}\!\left[\rAct_{\mathcal{C}},\,\rEnvF_{\mathcal{V}}\right]
  - E\!\left[\rho^{\mathrm{NH_3}}\right],
  \label{eq:eads}
\end{equation}
with $\mathcal{C}$ the promoted cluster, $\mathcal{V}$ its void, the first term relaxed in the
frozen potential, the second evaluated at the fixed cluster geometry in the same potential, and $E[\rho^{\mathrm{NH_3}}]$ the energy of the relaxed
gas-phase molecule.
The periodic reference is the same expression with the combined system in place of the
first term and the clean slab in place of the second.

\begin{figure}
  \centering
  \includegraphics[width=0.7\linewidth]{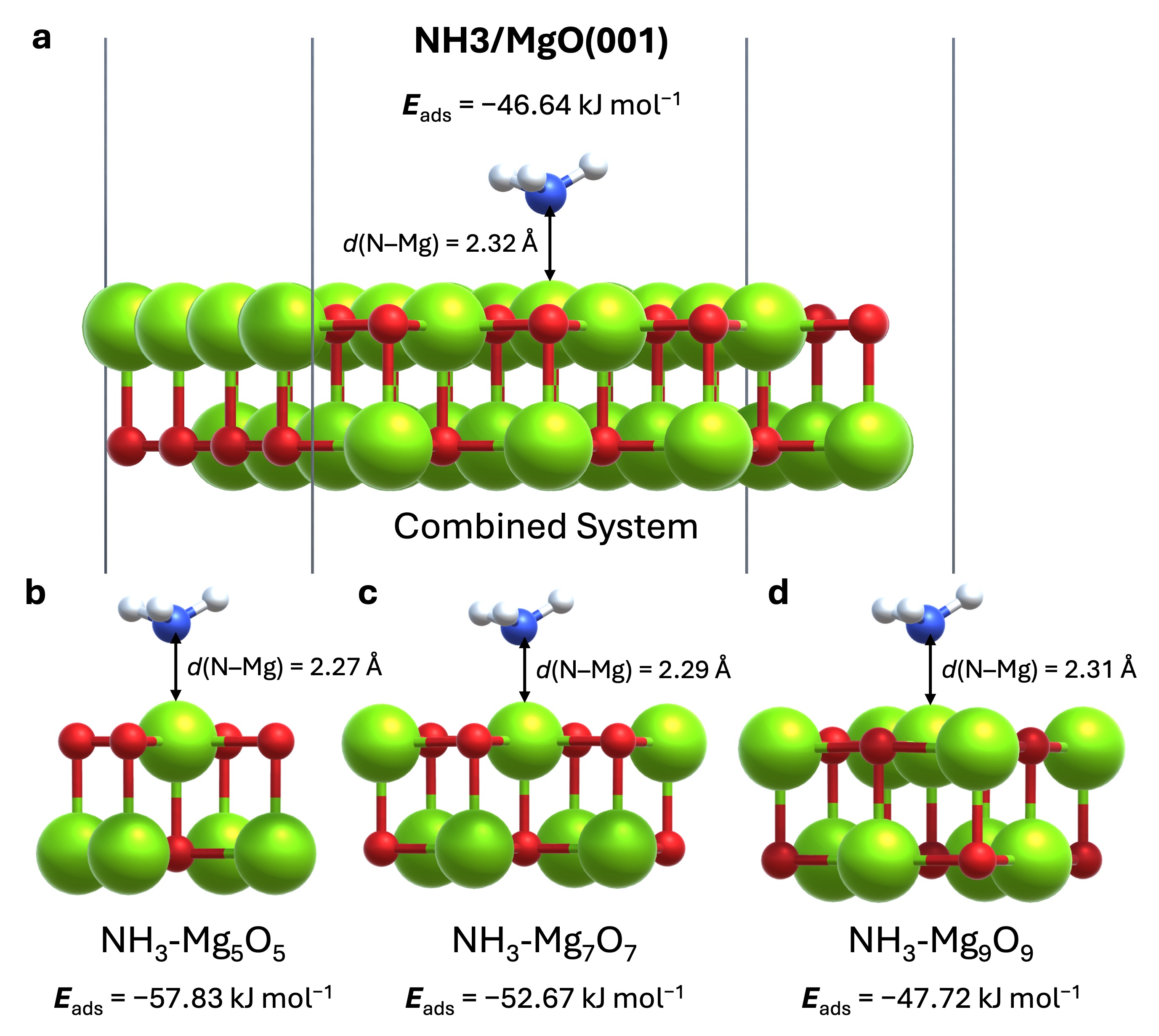}
  \caption{Optimized structures of NH$_3$ on MgO(001). \textbf{a}, The relaxed combined system of
  68 atoms in the two-dimensionally periodic surface cell. \textbf{b}--\textbf{d}, The three relaxed
  active subsystems NH$_3+$Mg$_5$O$_5$ (\textbf{b}), NH$_3+$Mg$_7$O$_7$ (\textbf{c}) and
  NH$_3+$Mg$_9$O$_9$ (\textbf{d}). Only the four atoms of NH$_3$ are free in each case, and every
  substrate atom is held at the reference geometry. Structures are relaxed at the PBE/def2-TZVP level. The adsorption energy $E_{\mathrm{ads}}$ and the
  N--Mg distance are given for each structure (Tables~\ref{tab:nh3-mgo} and~\ref{tab:nh3-mgo-geom}).}
  \label{fig:nh3-mgo-struct}
\end{figure}

Table~\ref{tab:nh3-mgo} reports the relaxed adsorption energies.
Convergence with the cluster size is monotonic, similar to CO.
The embedded adsorption energy deviates from the combined-system reference by $-11.20$,
$-6.03$, and $-1.08$~kJ/mol for NH$_3+$Mg$_5$O$_5$, NH$_3+$Mg$_7$O$_7$, and
NH$_3+$Mg$_9$O$_9$, so a 22-atom active subsystem reproduces a $-46.64$~kJ/mol binding
energy to 1.1~kJ/mol (well below chemical accuracy).
Embedding NH$_3$ alone, with no promoted host atoms at all, overbinds by 12.04~kJ/mol,
a quarter of the adsorption energy, and promoting the first Mg$_5$O$_5$ shell improves it by less
than 1 kJ/mol.

\begin{table}
  \centering
  \caption{Adsorption energy of NH$_3$ on MgO(001) from embedded active subsystems, NH$_3$ alone or with promoted MgO clusters of growing size, against the periodic calculation of the combined system. PBE/def2-TZVP, without counterpoise correction. Cycles are the geometry-optimization cycles to convergence.
  Deviations are embedded minus combined
  system.}
  \label{tab:nh3-mgo}
  \begin{tabular}{lrrrr}
    \toprule
    active subsystem & atoms & cycles & $E_{\mathrm{ads}}$/kJ mol$^{-1}$
    & $\Delta E_{\mathrm{ads}}$/kJ mol$^{-1}$ \\
    \midrule
    NH$_3$ only          &  4 & 21 & $-58.68$ & $-12.05$ \\
    NH$_3+$Mg$_5$O$_5$   & 14 &  8 & $-57.83$ & $-11.20$ \\
    NH$_3+$Mg$_7$O$_7$   & 18 &  9 & $-52.67$ & $-6.03$  \\
    NH$_3+$Mg$_9$O$_9$   & 22 & 10 & $-47.72$ & $-1.08$  \\
    \midrule
    combined system      & 68 &  8 & $-46.64$ & --       \\
    \bottomrule
  \end{tabular}
\end{table}

Next, we observe the geometrical properties of the relaxed adsorbate.
This is the more important test of the validity and accuracy of the analytical gradient implementation.
The relaxed structures in Table~\ref{tab:nh3-mgo-geom} show the same monotonicity of the averaged bond lengths and angles as observed for adsorption energies.
The averaged intramolecular coordinates $r(\mathrm{N}\text{-}\mathrm{H})$ and $\angle(\mathrm{H}\text{-}\mathrm{N}\text{-}\mathrm{H})$ are within 0.0007~\AA\ and 0.8$^{\circ}$ of the periodic reference for the bare NH$_3$ molecule as the active subsystem, and within 0.0001~\AA\ and 0.08$^{\circ}$ for the NH$_3+$Mg$_9$O$_9$ active subsystem.
Furthermore, the bare embedded molecule sits 0.164~\AA\ too close to the site Mg, and that error falls to 0.055, 0.036, and 0.015~\AA\ as the size of the cluster grows.
The averaged $\angle(\mathrm{Mg}\text{-}\mathrm{N}\text{-}\mathrm{H})$ follows the same sequence, from 110.26$^{\circ}$ for the bare molecule to 110.88$^{\circ}$ at NH$_3+$Mg$_9$O$_9$ against 110.99$^{\circ}$ for the combined system.
Figure~\ref{fig:nh3-mgo-struct} shows the corresponding optimized structures.
\begin{table}
  \centering
  \caption{Geometrical attributes of the relaxed structure of NH$_3$ on MgO(001) from embedded active subsystems of
  growing size, against the periodic calculation of the combined system (PBE/def2-TZVP, structures of Figure~\ref{fig:nh3-mgo-struct}).  $r(\mathrm{N}\text{-}\mathrm{H})$,
  $\angle(\mathrm{H}\text{-}\mathrm{N}\text{-}\mathrm{H})$ and
  $\angle(\mathrm{Mg}\text{-}\mathrm{N}\text{-}\mathrm{H})$ are
  averages over the three bonds and angles of the adsorbate, and Mg is the
  substrate magnesium nearest to N. For the bare embedded molecule, Mg belongs to the
  frozen host, so $d(\mathrm{N}\text{-}\mathrm{Mg})$ is measured across the subsystem boundary.}
  \label{tab:nh3-mgo-geom}
  \begin{tabular}{lrrrr}
    \toprule
    structure & $r(\mathrm{N}\text{-}\mathrm{H})$/\AA
    & $\angle(\mathrm{H}\text{-}\mathrm{N}\text{-}\mathrm{H})$/$^{\circ}$
    & $d(\mathrm{N}\text{-}\mathrm{Mg})$/\AA
    & $\angle(\mathrm{Mg}\text{-}\mathrm{N}\text{-}\mathrm{H})$/$^{\circ}$ \\
    \midrule
    NH$_3$ gas phase     & 1.0223 & 106.41 & --     & --     \\
    NH$_3$ only          & 1.0201 & 108.65 & 2.1591 & 110.26 \\
    NH$_3+$Mg$_5$O$_5$   & 1.0216 & 108.16 & 2.2678 & 110.70 \\
    NH$_3+$Mg$_7$O$_7$   & 1.0218 & 108.07 & 2.2875 & 110.76 \\
    NH$_3+$Mg$_9$O$_9$   & 1.0221 & 107.91 & 2.3082 & 110.88 \\
    \midrule
    combined system      & 1.0222 & 107.83 & 2.3231 & 110.99 \\
    \bottomrule
  \end{tabular}
\end{table}

Once the host density exists, every SCF iteration of an embedded run evaluates only
molecular integrals over the active subsystem.

\subsection{Solvatochromic shifts in excitation energy from RT-TDDFET}
\label{sec:rt-results}

The preceding sections established the accuracy of the direct-space embedding for
ground-state energetics and for analytic gradients (forces).
We now test its extension to RT-TDDFET as described in Section~\ref{sec:extensions}, i.e., whether the embedding potential, computed once and held fixed during the propagation, reproduces an excited-state observable of the active subsystem.
The observable chosen here is the solvatochromic shift of the lowest $\pi\rightarrow\pi^*$
excitation of MCP in water.
This is a prototypical system used in the embedding literature and has been shown to exhibit a modest blue-shift in the presence of water solvent.\cite{Daday.König.ea:State-Specific.2013,Daday.Konig.ea:Wavefunction.2014,Sharma.Sierka:Efficient.2022,Franzke2026RecentTurbomole}

The combined system contains an MCP molecule solvated in bulk water (64 water molecules).
The MCP molecule is considered the active subsystem and the bulk water molecules the host (environment).
The spectrum of the isolated molecule is computed using the same solvated geometry leaving out the water molecules, so that the shift observed is a solvent effect and not a geometry effect.

To see the effect of solvation on the lowest excitation energy, we compute the absorption spectra of the solvated MCP molecule using RT-TDDFET with the frozen embedding potential.
It is built from the fitted density of the clean periodic bulk water host alone, with the solute sitting in the cavity of that host, so that no host sites are removed (no-void branch of Figure~\ref{fig:scheme}b).
Additionally, we also compare the results with our older molecule-in-periodic embedding implementation of ref~\citenum{Sharma.Sierka:Efficient.2022}, denoted \emph{previous work}, in which the environment density is approximated by subtraction, $\rEnvF = \rTot - \rAct_{\mathrm{iso}}$, using the converged density of the combined periodic system.

Additionally, our recent implementation of periodic LR-TDDFT in \textsc{riper}, to be described separately, is used to compute the lowest excitation energy of the combined periodic system (200 atoms).
This enables a direct comparison of the embedded spectra with an actual reference value.

The active subsystem is propagated with either PBE or the SCAN0 hybrid meta-GGA functional,\cite{Hui.Chai:SCAN-based.2016} and the whole comparison is repeated for the pob-DZVP-rev2 and pob-TZVP-rev2 basis sets~\cite{VilelaOliveira.Laun.ea:BSSE.2019} to investigate their effect.

\begin{figure}
  \centering
  \includegraphics[width=\linewidth]{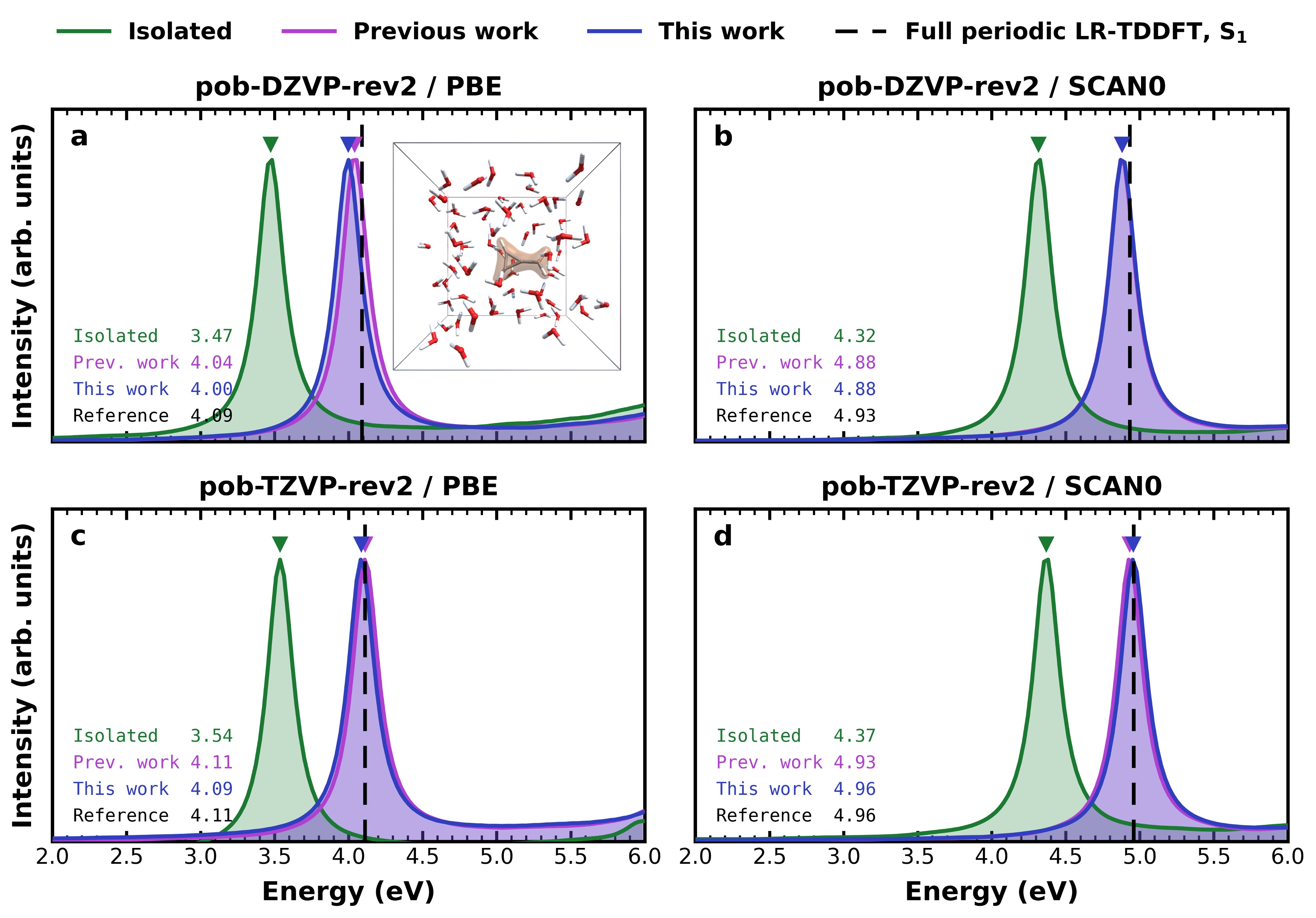}
  \caption{Absorption spectra of MCP in water from RT-TDDFET. \textbf{a}, pob-DZVP-rev2 with PBE;
  \textbf{b}, pob-DZVP-rev2 with SCAN0; \textbf{c}, pob-TZVP-rev2 with PBE; \textbf{d}, pob-TZVP-rev2
  with SCAN0, where the functional is that of the active-subsystem propagation and the basis set is
  used throughout. Green: the isolated molecule at the solvated geometry, from RT-TDDFT. Blue: the
  solute embedded in the direct-space potential of this work. Magenta: the solute embedded in the
  subtractive potential of ref~\citenum{Sharma.Sierka:Efficient.2022}. Embedding potentials at the PBE/LC94 level. The dashed line is the lowest
  excitation of the complete 200-atom periodic system from LR-TDDFT with the functional and basis of
  the panel. The numbers in each panel are the peak positions in eV. The inset in \textbf{a} shows the
  periodic cell of MCP and 64 water molecules.}
  \label{fig:mcp-water}
\end{figure}

Figure~\ref{fig:mcp-water} shows the resulting spectra and the peak positions from various calculations.
The isolated molecule absorbs at 3.47~eV with PBE and at 4.32~eV with SCAN0 in the double-zeta
basis, and at 3.54 and 4.37~eV respectively in the triple-zeta basis.
The combined-system reference places the same band at 4.09 and 4.93~eV (pob-DZVP-rev2) and at 4.11 and 4.96~eV (pob-TZVP-rev2).
The solvent shift to be reproduced is therefore between 0.57 and 0.62~eV, depending on basis and functional.
In the triple-zeta basis, the direct-space potential lands 0.02~eV below the reference with PBE
and exactly on it with SCAN0.
The spectra from the subtractive embedding scheme perform similarly, landing very close to the reference.
In the double-zeta basis the deviations grow to 0.09 and 0.05~eV for the direct-space
potential and to 0.05 and 0.05~eV for the subtractive one.
Improving the basis therefore improves the embedding accuracy and it reproduces a 0.6~eV solvent shift (reference) to better than 0.02~eV.
The direct-space potential matches a construction that requires the combined-system calculation,
while never performing one.

An interesting observation can be made by looking at the functional dependence.
Changing the active subsystem from PBE to SCAN0 moves the absolute peak position by about 0.85~eV, which is the well-known effect of exact exchange on a valence $\pi\rightarrow\pi^*$ transition.
However, it barely moves the predicted solvent shift by at most 0.04~eV, indicating minimal dependence of the solvatochromic shift on the functional.
This could explain why, despite the embedding potential being built at the PBE/LC94 level in every case, it reproduces the shift almost independently of the functional used for the active subsystem.
This demonstrates the advantage of a high-level-in-low-level treatment.
One pays for the better functional only on the 8 atoms of the solute, and still obtains the environment response that a full calculation with that functional would give.

\subsection{High harmonic generation spectra of adsorbates using RT-TDDFET}
\label{sec:hhg-results}

The absorption spectra study in the previous section investigated the linear response of the active subsystem to a weak perturbation.
In this section, we investigate the effect of the substrate on the adsorbate for a nonlinear process like HHG by calculating its HHG spectrum under the influence of a strong laser pulse.
In HHG, an intense laser field drives the electrons of a molecule far from their ground-state distribution, and the emitted radiation contains high harmonics of the driving frequency.
In the semiclassical three-step model of Corkum,\cite{Corkum:Plasma.1993} an electron tunnels out, is accelerated by the field, and recombines with the parent ion.
This produces a plateau of harmonics of similar intensity that ends at the cutoff energy $E_{\mathrm{c}} = I_{\mathrm{p}} + 3.17\,U_{\mathrm{p}}$,\cite{Krause.Schafer.ea:High-order.1992,Corkum:Plasma.1993} where $I_{\mathrm{p}}$ is the ionization potential and $U_{\mathrm{p}} = E_0^2/(4\omega_0^2)$ the ponderomotive energy of an electron in a field of amplitude $E_0$ and frequency $\omega_0$ (atomic units).
Since the driven electron traverses the region around the molecule, HHG from an adsorbate is expected to be strongly affected by the substrate.
As a test case, we consider CO$_2$ physisorbed on an hBN monolayer.

To obtain the lowest-energy adsorption geometry, we started from eight flat configurations of CO$_2$ on hBN monolayer, with the carbon atom above a B atom, an N atom, the center of a hexagon, or a B-N bond and the molecular axis along the zigzag or the armchair direction, and from one configuration with the molecule perpendicular to the sheet.
The most stable structure has CO$_2$ lying flat, with C above N and the molecular axis along the armchair direction, 3.12~\AA{} above the sheet, and an adsorption energy of $-21.9$~kJ/mol.
All flat configurations lie within 0.7~kJ/mol of this structure, while the perpendicular one is 10.5~kJ/mol less stable.
The adsorbed molecule keeps its gas-phase bond length of 1.181~\AA{} and is bent by only 0.7$^{\circ}$.

For the embedding calculations, CO$_2$ is the active subsystem and the relaxed hBN sheet of this structure, without the molecule, is the periodic host (no-void branch of Figure~\ref{fig:scheme}b).

The spectrum of the isolated molecule is computed for CO$_2$ at its optimized gas-phase geometry.

The active subsystem is driven by a linearly polarized laser pulse,
\begin{equation}
  \mathbf{E}(t) = E_0\,\hat{\mathbf{z}}\,\sin(\omega_0 t)\cos^2\!\left[\frac{\pi(t-\sigma)}{2\sigma}\right],
  \qquad 0 \le t \le 2\sigma,
  \label{eq:hhg-field}
\end{equation}
with a wavelength of 800~nm ($\hbar\omega_0 = 1.55$~eV), a peak intensity of $I_0 = 1\times10^{14}$~W/cm$^2$, which corresponds to $E_0 = 0.0534$~a.u.\ through $I_0 = \frac{1}{2}\varepsilon_0 c E_0^2$, and $\sigma = 1500$~a.u., the full width at half maximum of the field envelope, so that the pulse lasts $2\sigma = 72.6$~fs or 27 optical cycles.
The field is polarized along the surface normal $\hat{\mathbf{z}}$, i.e., perpendicular to the molecular axis, and the isolated molecule is oriented in the same way.
The field couples to the electrons in the length gauge within the dipole approximation, i.e., the interaction $-\hat{\bm{\mu}}\cdot\mathbf{E}(t)$ is added to the KS matrix at every time step, together with the embedding potential, which is held fixed at its ground-state form.
The KS equations are integrated with the second-order Magnus propagator\cite{Magnus:On.1954,Blanes.Casas.ea:Magnus.2009,Castro.Marques.ea:Propagators.2004} with a self-consistent predictor-corrector step\cite{Mueller.Sharma.ea:Real-time.2020} and a time step of 0.1~a.u.\ over the whole pulse (30\,000 steps).

Following our previous work,\cite{Sharma.Franzke.ea:Density.2025} the HHG power spectrum $P(\omega)$ is computed from the dipole acceleration,
\begin{equation}
  \begin{split}
  P(\omega) &= \sum_{j}\left|\frac{1}{t_f-t_i}\int_{t_i}^{t_f} W(t)\,
  \frac{d^2\mu_j^{\mathrm{ind}}(t)}{dt^2}\,e^{-i\omega t}\,dt\right|^2 \\
  &\approx \frac{\omega^4}{(t_f-t_i)^2}\sum_{j}\left|\int_{t_i}^{t_f} W(t)\,
  \mu_j^{\mathrm{ind}}(t)\,e^{-i\omega t}\,dt\right|^2,
  \end{split}
  \label{eq:hhg-power}
\end{equation}
where $W(t) = \sin^2[\pi(t-t_i)/(t_f-t_i)]$ is the Hann window over the propagation interval $[t_i, t_f] = [0, 2\sigma]$ and
\begin{equation}
  \mu_j^{\mathrm{ind}}(t) = \mu_j(t) - \mu_j^0, \qquad
  \mu_j(t) = \Tr\!\left[\mathbf{M}^j \matD(t)\right]
  \label{eq:hhg-dipole}
\end{equation}
with $\mu_j(t)$ the dipole moment along $j = x, y, z$, $\mu_j^0$ its value before the pulse, $\mu_j^{\mathrm{ind}}(t)$ the induced dipole moment, $\mathbf{M}^j$ the dipole moment matrix and $\matD(t)$ the time-dependent density matrix of the active subsystem.
Instead of differentiating the dipole moment numerically, we evaluate the acceleration form in the frequency domain, where the second time derivative becomes a factor of $-\omega^2$ when the derivatives of the window are neglected (second expression of eq~\ref{eq:hhg-power}).

\begin{figure}[!ht]
  \centering
  \includegraphics[width=0.8\linewidth]{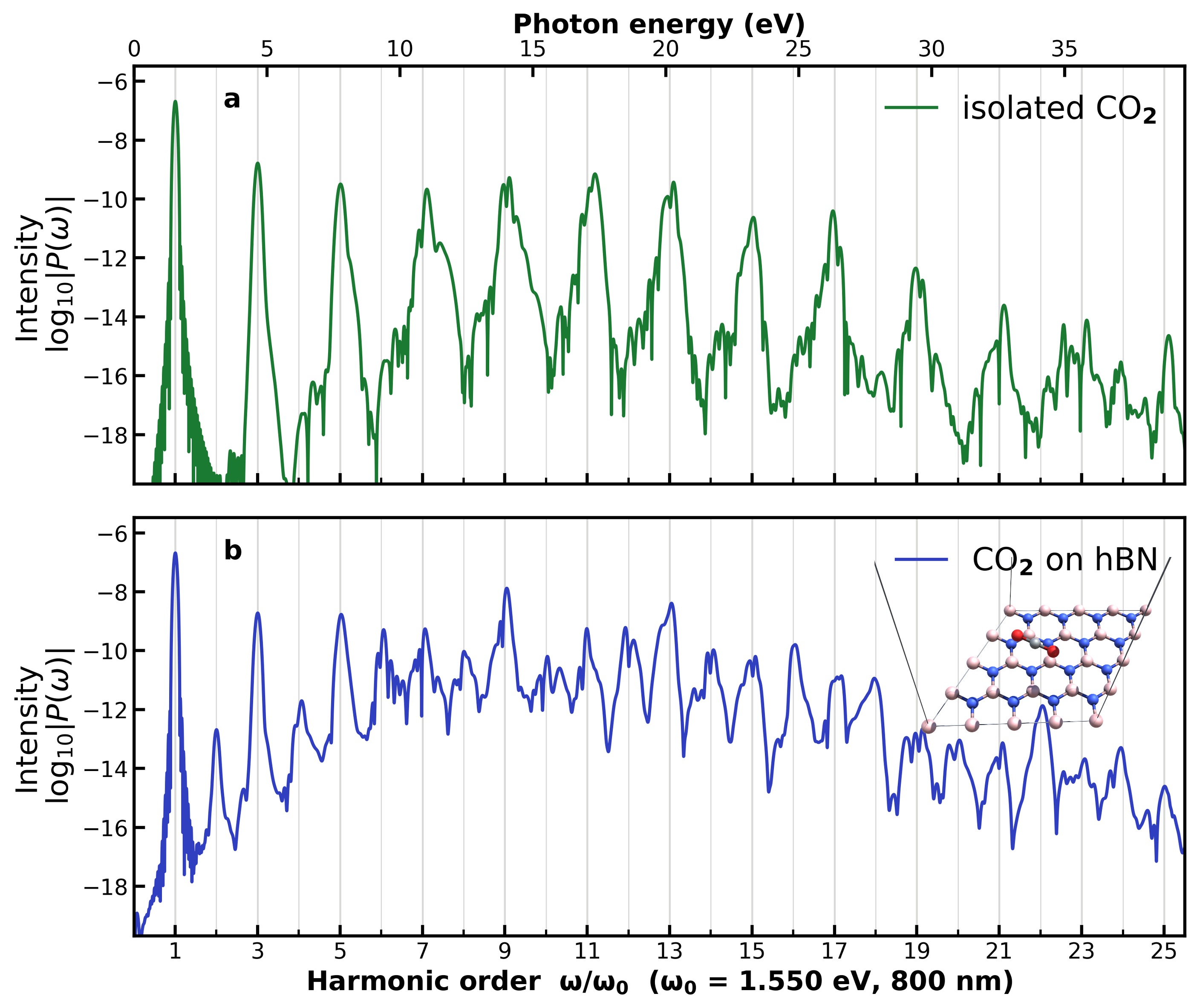}
  \caption{HHG spectra of CO$_2$, $\log_{10}|P(\omega)|$ of eq~\ref{eq:hhg-power} against the harmonic order, for an 800~nm pulse of $1\times10^{14}$~W/cm$^2$ (PBE/def2-TZVPD). \textbf{a}, The isolated molecule at its gas-phase geometry,
  from RT-TDDFT. \textbf{b}, The molecule adsorbed on the hBN monolayer, from RT-TDDFET in the frozen
  embedding potential of the periodic host, with the field along the surface normal. The inset shows
  the adsorption structure in the $4\times4$ hBN cell.}
  \label{fig:co2-hbn-hhg}
\end{figure}

Figure~\ref{fig:co2-hbn-hhg} shows the resulting spectra.
The isolated molecule (panel a) displays the textbook shape of an HHG spectrum.
The intensity drops by two orders of magnitude from the fundamental to the third harmonic (perturbative region), stays within a factor of about three from the 5th to the 13th harmonic (plateau), decreases by an order of magnitude at the 15th and 17th harmonics, and drops by two further orders of magnitude at the 19th harmonic (cutoff).
Only odd harmonics appear, as expected from the inversion symmetry of CO$_2$.
With $U_{\mathrm{p}} = 5.98$~eV and $I_{\mathrm{p}}$ approximated by the negative PBE highest occupied molecular orbital (HOMO) energy of 9.07~eV, the three-step model predicts a cutoff of 28.0~eV, i.e., at the 18th harmonic, in close agreement with the computed spectrum.

The spectrum of the adsorbed molecule (panel b) retains the same overall structure, but the substrate affects it in two ways.
Firstly, most of the plateau harmonics are enhanced, the 9th and 13th harmonics by factors of 25 and 11, while the fundamental and the third harmonic change by at most 15\%.
The cutoff stays at the 19th harmonic, consistent with a HOMO energy that is lowered by only 14~meV.
Secondly, even harmonics appear, and in the plateau the 6th, 12th, and 16th harmonics become as intense as their odd neighbors.
Their origin is expected from the broken inversion symmetry of the adsorbate in the presence of the substrate.
The field drives the electron alternately towards the hBN sheet, where it experiences the electrostatic potential of the host and the repulsive nonadditive kinetic potential, and away from it into vacuum, so that the response in the two half cycles is no longer equivalent.

After the pulse, the absorbed energy is only 0.14 and 1.5~meV for the isolated and the embedded molecule, respectively, so that both return essentially to their ground state.

The embedded propagation took 32~min on 16 cores, compared to 29~min for the isolated molecule.
Because the host density is frozen, the dynamic polarization of hBN and electron transfer into the substrate are not included.
\subsection{Solvatochromic shift in excitation energy from CC2-in-DFT}
\label{sec:cc2-results}

The RT-TDDFET study in Section~\ref{sec:rt-results} showed that the frozen embedding potential reproduces the solvent shift of an excitation energy when the active subsystem is described with DFT.
In this section, we test the WFT-in-DFT extension of Section~\ref{sec:extensions}, in which the exported embedding potential is added to the Hartree-Fock core Hamiltonian of the active subsystem and the excitation energy is calculated with the correlated CC2 method.
We again consider the lowest $\pi\rightarrow\pi^*$ excitation of MCP in bulk water, using the same periodic cell of one MCP and 64 water molecules as in Section~\ref{sec:rt-results}.
The MCP molecule is the active subsystem and the bulk water is the periodic host.
The excitation energy of the isolated molecule is calculated at the same solvated geometry, so that the computed shift is again a pure solvent effect.

\begin{figure}[!h]
  \centering
  \includegraphics[width=0.8\linewidth]{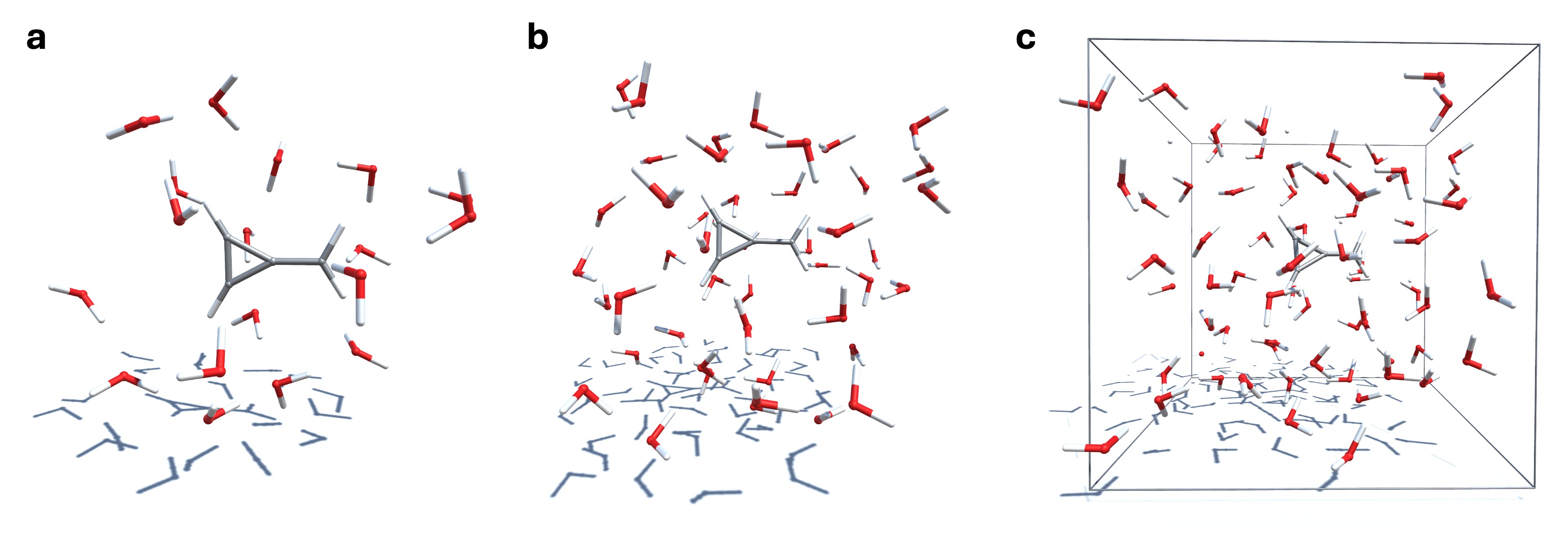}
\caption{MCP in water. \textbf{a}, MCP with its first hydration shell (18 water molecules) and
  \textbf{b}, MCP with its first and second hydration shells (45 water molecules), both carved out
  of the periodic cell and used for the supermolecular CC2 reference calculations. \textbf{c}, The
  periodic cell of one MCP and 64 water molecules (cubic, $a = 12.610$~\AA).}
  \label{fig:mcp-cc2-struct}
\end{figure}

Unlike the RT-TDDFET case, a periodic CC2 calculation of the combined 200-atom system is not available as a reference.
Instead, the reference values are obtained from supermolecular CC2 calculations of finite clusters of MCP and water molecules carved out of the periodic cell.
The water molecules are assigned to hydration shells according to their hydrogen-bond connectivity.
The first shell consists of the 18 water molecules in direct contact with MCP, each with at least one atom within 3.16~\AA{} of an MCP atom, while the next water molecule approaches MCP no closer than 3.56~\AA.
The second shell consists of the 27 water molecules outside the first shell whose oxygen atom lies within 3.5~\AA{} of the oxygen atom of a first-shell water molecule, evaluated with the minimum-image convention.
Two finite clusters with complete shells are thus considered: MCP$+$single-shell with 18 water molecules and MCP$+$double-shell with 45 water molecules.
Figure~\ref{fig:mcp-cc2-struct} shows both clusters together with the periodic cell.

Table~\ref{tab:mcp-cc2} reports the resulting excitation energies.
The isolated molecule has the lowest excitation energy at 4.36~eV.
In the periodic embedding potential, the CC2-in-DFT calculation predicts the same excitation at 4.97~eV, a blue shift of 0.61~eV.
The supermolecular CC2 calculations give 4.96 and 4.98~eV for MCP$+$single-shell and MCP$+$double-shell, corresponding to shifts of 0.60 and 0.62~eV, respectively.
The periodic CC2-in-DFT result thus lies within 0.01~eV of both.
The CC2-in-DFT shift also falls within the range of 0.57 to 0.62~eV obtained from TDDFT in Section~\ref{sec:rt-results}, which again indicates that the solvent shift of this transition depends little on the method used for the active subsystem.

\begin{table}
  \centering
  \caption{Lowest singlet excitation energy $E_1$ of MCP in water from CC2 and its solvent shift $\Delta E_1$ with respect to the isolated molecule at the same geometry. RI-CC2/def2-TZVPD for MCP and pob-TZVP-rev2 for water, embedding potentials at the PBE/LC94 level. In the cluster rows $E_1$ is the supermolecular CC2 value, in the last row the periodic CC2-in-DFT value. 
  CC2-in-DFT$_{\mathrm{cl}}$ denotes MCP embedded in a potential whose molecular host consists of the water molecules of the same cluster. 
  $\Delta_{\mathrm{emb}}$ is CC2-in-DFT$_{\mathrm{cl}}$ minus supermolecular CC2, and $\Delta_{\mathrm{per}}$ is supermolecular CC2 minus periodic CC2-in-DFT. 
  All energies are in eV, and differences are computed from unrounded values.}
  \label{tab:mcp-cc2}
  \begin{tabular}{lccccccc}
    \toprule
    system & waters & basis fn. & $E_1$ & $\Delta E_1$ & CC2-in-DFT$_{\mathrm{cl}}$
    & $\Delta_{\mathrm{emb}}$ & $\Delta_{\mathrm{per}}$ \\
    \midrule
    isolated MCP        &  0 &  184 & 4.36 & --       & --    & --       & --       \\
    MCP$+$single-shell       & 18 &  724 & 4.96 & $+0.60$ & 5.01 & $+0.05$ & $-0.01$ \\
    MCP$+$double-shell  & 45 & 1534 & 4.98 & $+0.62$ & 5.02 & $+0.04$ & $+0.01$ \\
    \midrule
    periodic CC2-in-DFT & 64 &  184 & 4.97 & $+0.61$ & --    & --       & --       \\
    \bottomrule
  \end{tabular}
\end{table}

The construction of the finite reference clusters requires care.
Clusters that are cut by distance in the middle of the second shell, with 26 and 35 water molecules, yield shifts of only 0.36 and 0.29~eV, i.e., 0.25 and 0.32~eV below the periodic CC2-in-DFT value.
The lowest excitation of MCP is accompanied by a large change of the dipole moment and is therefore sensitive to the electric field of the surrounding water molecules.
In the bulk, the orientations of the water molecules of a shell are balanced by their neighbors, while a partially filled shell leaves a net, uncompensated field at the solute.
As a consequence, the excitation energy of finite clusters does not converge monotonically with the number of water molecules, and only clusters made of complete shells, or the periodic environment itself, provide a meaningful reference.

To separate the error of the embedding itself from the effect of the size of the environment, the CC2-in-DFT calculations were repeated with embedding potentials built from exactly the water molecules of each finite cluster as a molecular host.
For the same environment, the embedded excitation energy lies only 0.05 and 0.04~eV above the supermolecular value for MCP$+$single-shell and MCP$+$double-shell, respectively.
This small residual is consistent with the state-independent approximation of Section~\ref{sec:extensions}: the frozen environment cannot respond to the excitation, whereas the explicit water electrons in the supermolecular calculation contribute their polarization and dispersion to the CC2 description of the excited state.
The water molecules beyond the second shell, which are present only in the periodic host, lower the embedded excitation energy by a further 0.05~eV (5.02 against 4.97~eV).

The periodic host calculation took 5.0~min on 32 threads, and the embedded DFT calculation of MCP that exports the embedding potential took 41~s.
The subsequent Hartree-Fock and CC2 calculations of the embedded molecule, with its 184 basis functions, took 35~s on 8 threads, essentially the cost of the isolated molecule (31~s).
In contrast, the supermolecular CC2 calculation took 1.6~h on 24 threads for MCP$+$single-shell (724 basis functions) and 27.5~h on 64 threads for MCP$+$double-shell (1534 basis functions).
In terms of CPU time, the complete CC2-in-DFT workflow, including the one-time host calculation, required 3.1~CPU hours, compared to 1760~CPU hours for the largest cluster.
\section{Discussion}
\label{sec:conclusions}

We have formulated and implemented a direct-space DFT embedding framework for molecular subsystems in molecular and periodic environments within \textsc{riper}. The environment is included in the embedded calculation through its fitted electron density, auxiliary basis, nuclear charges, and lattice, while the long-range electrostatic interactions are evaluated by combining an explicit crystal near-field with the CFMM far-field operator. The same construction applies from finite molecular hosts to periodic systems of arbitrary dimensionality. Analytic nuclear gradients of the complete embedding functional further make structural optimization of the active subsystem possible in the presence of the frozen environment.

The adsorption studies on MgO(001) provide a direct assessment against calculations of the complete periodic systems and establish the size of the active subsystem as a systematic convergence parameter. For CO on MgO(001), increasing the promoted MgO cluster gives rapid and monotonic convergence of the interaction-energy curve. With CO$+$Mg$_9$O$_9$ as the 20-atom active subsystem, the interaction energy differs from the 66-atom periodic reference by only 0.44~kJ/mol at the minimum, with a root-mean-square deviation of 0.50~kJ/mol over the complete 15-point curve. At the same time, the full embedded scan, including the one-time host calculation, is 27 times less expensive in CPU time than evaluating the corresponding periodic combined systems. The NH$_3$/MgO(001) optimizations demonstrate that the analytic gradients provide the same systematic convergence for relaxed structures. A 22-atom NH$_3+$Mg$_9$O$_9$ active subsystem reproduces the periodic adsorption energy within 1.08~kJ/mol and the N--Mg adsorption distance within 0.015~\AA, while the intramolecular geometry agrees still more closely. These results show that promoting a sufficiently large local region moves the subsystem boundary away from the chemically important interaction region and allows both energetics and structures to approach the periodic reference systematically.

The same embedding potential can be reused without modification for time-dependent calculations. For methylenecyclopropene in periodic water, RT-TDDFET reproduces the approximately 0.6~eV solvatochromic blue shift of the complete 200-atom periodic system. With the triple-zeta basis, the embedded excitation differs from the periodic LR-TDDFT reference by only 0.02~eV with PBE and coincides with it for SCAN0, despite the embedding potential itself being constructed at the PBE/LC94 level. Changing the treatment of the active subsystem shifts the absolute excitation energy substantially but changes the solvent shift only weakly, illustrating the utility of describing a small active region at a higher level while retaining the extended environment through a low-cost frozen potential.

The strong-field calculations of CO$_2$ on hBN further show that the approach is not limited to linear-response properties. The static potential of the substrate qualitatively changes the HHG spectrum of the adsorbate: even harmonics appear because the environment breaks the inversion symmetry of isolated CO$_2$, while selected plateau harmonics are enhanced by more than an order of magnitude, by factors of 25 and 11 for the ninth and thirteenth harmonics, respectively. Nevertheless, the embedded propagation requires 32~min compared with 29~min for the isolated molecule. Thus, once the host density has been generated, the effect of an extended surface on strong-field molecular electron dynamics can be studied at essentially the computational cost of a gas-phase propagation.

Finally, exporting the converged potential as a one-electron matrix extends the same construction beyond DFT. CC2-in-DFT predicts the lowest excitation of MCP in periodic water at 4.97~eV, corresponding to a 0.61~eV solvent blue shift. This lies within 0.01~eV of supermolecular CC2 calculations containing complete first and first-plus-second hydration shells. The comparison also emphasizes an advantage of the periodic description: finite clusters cut through incomplete hydration shells produce large and nonmonotonic errors because of uncompensated solvent fields. The periodic embedding treatment retains the complete bulk environment while the correlated calculation itself contains only the 184 basis functions of MCP. Including the one-time periodic host calculation, the complete CC2-in-DFT workflow requires 3.1~CPU hours, compared with 1760~CPU hours for the largest supermolecular CC2 reference calculation.

The frozen environment remains the central approximation of the present formulation. Its electronic density does not respond dynamically to the active subsystem, so differential environment polarization, charge transfer into the host, and related response effects are absent; the CO$_2$/hBN HHG calculation, for example, isolates the influence of the static substrate potential rather than the full coupled electron dynamics of adsorbate and surface. Within this approximation, however, the present results show that one reusable host density can support ground-state energetics, structural relaxation, linear and nonlinear time-dependent response, and correlated wave-function calculations while avoiding repeated periodic calculations of the combined system. Extensions that introduce environment response and analytic gradients for the WFT-in-DFT route provide natural directions for further development. Moreover, the combination of analytic gradients with a cost per energy and gradient evaluation comparable to that of a gas-phase calculation makes the present scheme well suited for applications that require a large number of such evaluations. Examples are the exploration of reaction pathways and barriers of adsorbates on surfaces with the nudged elastic band method\cite{Henkelman.Uberuaga.ea:climbing.2000} and ab initio molecular dynamics of molecules on surfaces and in solvents or crystals. For processes that are accompanied by a substantial rearrangement of the host, such applications could be combined with an update of the host density along the reaction path or trajectory.

\section{Methods}
\label{sec:comp}

The calculations are performed using a development version of TURBOMOLE\cite{Franzke.Holzer.ea:TURBOMOLE.2023} and the \textsc{riper} module.\cite{Sharma.Franzke.ea:Density.2025,Lazarski.Burow.ea:Density.2015,Lazarski.Burow.ea:Density.2016}
Structures were visualized with CrysX-3D Viewer.\cite{Sharma.Mishra:CrysX.2019}

\subsection{CO on MgO(001)}
The structure used here is the two-layer $2\times2$ slab of ref~\citenum{Ye.Berkelbach:Adsorption.2024}, taken from the supporting-data repository of that work, with CO being C-down on a five-fold coordinated surface Mg. 
The combined system consists of 66 atoms in total: CO (2 atoms) and 64 atoms in a slab with cell length of $a = 11.944$~\AA.
The clean slab calculations for embedding and the total DFT calculations for reference use a $3\times3$ $\mathbf{k}$-point mesh.
For embedding calculations, the selection radius is $R = 6$~bohr, and the explicit near-field is $k = 2$.
All calculations employed the PBE exchange-correlation functional,\cite{Perdew.Burke.ea:Generalized.1996} the def2-TZVP orbital basis,\cite{Weigend.Ahlrichs:Balanced.2005} the universal Coulomb fitting basis,\cite{Weigend:Accurate.2006} quadrature grid m5,\cite{Treutler.Ahlrichs:Efficient.1995,Eichkorn.Weigend.ea:Auxiliary.1997} and the LC94 KEDF\cite{Lembarki.Chermette:Obtaining.1994} for embedding potential generation.

All calculations were run on a single node with one AMD Ryzen Threadripper PRO 5995WX processor (64 physical cores, 128 hardware threads, 4.6~GHz maximum clock) and 499~GB of memory.
The embedded active subsystems used 16 threads and the periodic calculations of the host and of the combined systems used 32. 
Because the two sets therefore differ in thread count, and because several runs shared the node, all costs are quoted as CPU times rather than wall times.

\subsection{NH$_3$ on MgO(001)}
The MgO(001) slab structure is the same as the one used for CO on MgO(001).
Other details like the basis sets, exchange correlation functional, KEDF, and $k$-point mesh are also the same.
For the analytical gradients (forces) the quadrature derivatives were also calculated.
The initial and the relaxed geometries are provided in the Supplementary Information. 

\subsection{MCP solvated in bulk water}
The unit cell used for this study consists of one MCP molecule solvated in bulk water (64 water molecules), 200 atoms in total, in a cubic cell of length 12.610~\AA{}.
The geometry was obtained from one of our previous publications~\cite{Franzke2026RecentTurbomole} and is also provided in the Supplementary Information.
The embedding potential of the current implementation is constructed from the periodic water host at the $\Gamma$-point with the PBE exchange-correlation functional\cite{Perdew.Burke.ea:Generalized.1996} and the LC94 KEDF,\cite{Lembarki.Chermette:Obtaining.1994} in the monomolecular basis, and the selection radius is $R = 10$~bohr. The potential of the previous implementation is obtained by subtraction from the converged density of the combined periodic system (Section~\ref{sec:rt-results}) with the same functional and KEDF.
PBE or the SCAN0 hybrid meta-GGA functional\cite{Hui.Chai:SCAN-based.2016} is used for the active subsystem density.
The pob-DZVP-rev2 and pob-TZVP-rev2 basis sets~\cite{VilelaOliveira.Laun.ea:BSSE.2019} are utilized for the whole system.

For the CC2-in-DFT calculations, the active subsystem (MCP) uses the def2-TZVPD basis set,\cite{Rappoport.Furche:Property-optimized.2010} while the water host uses the pob-TZVP-rev2 basis set.
The exported potential is added to the Hartree-Fock core Hamiltonian of MCP, and the lowest singlet excitation energy is calculated with the resolution-of-the-identity CC2 method (RI-CC2),\cite{Christiansen.Koch.ea:second-order.1995,Hattig.Weigend:CC2.2000} correlating all electrons.
The def2-TZVP auxiliary basis set\cite{Weigend.Haser.ea:RI-MP2.1998} is used on all atoms for the RI approximation in the CC2 calculations.
It is the recommended auxiliary basis for def2-TZVPD,\cite{Hellweg.Rappoport:Development.2015} and is also applied to the pob-TZVP-rev2 basis set of water, which is derived from def2-TZVP.
The supermolecular CC2 calculations of finite MCP-water clusters use the same basis set for each fragment as the embedding calculations, def2-TZVPD for MCP and pob-TZVP-rev2 for water.
The CC2 and CC2-in-DFT calculations were run on a single node with one AMD Ryzen Threadripper 3990X processor (64 physical cores, 128 hardware threads, 4.4~GHz maximum clock) and 256~GB of memory.

\subsection{CO$_2$ on hBN}
The CO$_2$ molecule was placed on a $4\times4$ supercell of the hBN monolayer and the geometry was optimized at the PBE-D3(BJ)/pob-TZVP-rev2 level,\cite{Perdew.Burke.ea:Generalized.1996,Grimme.Antony.ea:consistent.2010,Grimme.Ehrlich.ea:Effect.2011,VilelaOliveira.Laun.ea:BSSE.2019} with the lattice constant of the optimized primitive cell ($a = 2.510$~\AA), a $3\times3$ $\mathbf{k}$-point mesh, and all atoms relaxed within the fixed cell.
For the embedding calculations, the host (hBN) uses PBE, the pob-TZVP-rev2 basis, and the $3\times3$ $\mathbf{k}$-point mesh.
The active subsystem uses PBE and the def2-TZVPD basis,\cite{Rappoport.Furche:Property-optimized.2010} whose diffuse functions are required to describe the field-driven electron, especially for strong laser intensities.
The embedding potential is constructed with the PBE exchange-correlation functional and the LC94 KEDF,\cite{Lembarki.Chermette:Obtaining.1994} the selection radius is $R = 10$~bohr, and the explicit near-field is $k = 2$.

\section*{Data availability}
Cartesian coordinates of all
structures used in this work (CO and NH$_3$ on MgO(001), including the initial and relaxed embedded and periodic geometries and the promoted MgO clusters; 
MCP in periodic water and the carved MCP--water clusters with 18 and 45 water molecules; 
CO$_2$ on hBN) are provided as XYZ and extended XYZ files (the latter including lattice vectors for periodic systems) in a ZIP archive (Supplementary Data~1).
The absolute energies underlying the interaction energies of CO on MgO(001), the adsorption energies of NH$_3$ on MgO(001), and the adsorption energies of CO$_2$ on hBN are provided as a spreadsheet (Supplementary Data~2).
The time-dependent dipole moments and the absorption and high-harmonic spectra underlying Figures~\ref{fig:mcp-water} and~\ref{fig:co2-hbn-hhg}, together with the LR-TDDFT reference excitation energies and the laser-pulse parameters, are provided as plain-text files in a ZIP archive (Supplementary Data~3).
Supplementary Data~1 to~3 and all other datasets generated and analysed during the current study are available in the Zenodo repository, \url{https://doi.org/10.5281/zenodo.23953157}.

\section*{Code availability}
The underlying code for this study is not publicly available for proprietary reasons. The embedding scheme is implemented in the TURBOMOLE program package,\cite{Franzke.Holzer.ea:TURBOMOLE.2023} which is available from TURBOMOLE GmbH (\url{https://www.turbomole.org}); an educational version is available free of charge.

\bibliographystyle{naturemag-npj}
\bibliography{manuscript}

\section*{Acknowledgements}
M.~Sharma acknowledges the Anusandhan National Research Foundation for the National Post-Doctoral Fellowship (Grant No. PDF/2026/006571).
M.~Sharma also thanks TURBOMOLE GmbH for financial support.
M.~Sierka gratefully acknowledges financial support from DFG within the Collaborative Research Centre ``NOA -- Nonlinear Optics down to Atomic scales'' (CRC 1375 project no.\ 398816777, project A4) and the Research Training Group ``PhInt -- Photo-polarizable Interfaces and Membranes'' (GRK 3014 project no.\ 521747072), and from TURBOMOLE GmbH.

\section*{Author contributions}
M.~Sharma contributed to the implementation and execution of the study, the analysis of the results, and the writing of the manuscript.
M.~Sierka contributed to the implementation, analysis of the results, writing of the manuscript, and conceptual guidance and supervision.

\section*{Competing interests}
Marek Sierka has an equity interest in TURBOMOLE GmbH and serves as its chief executive officer.
Manas Sharma declares no
financial or non-financial competing interests.

\end{document}